    \documentclass[journal,article,submit,pdftex,moreauthors]{Definitions/mdpi} 
    \firstpage{1} 
    \pubvolume{1}
    \issuenum{1}
    \articlenumber{0}
    \usepackage{tikz}
    \usetikzlibrary{arrows.meta,positioning}

    \pubyear{2026}
    \copyrightyear{2026}
    \datereceived{ } 
    \daterevised{ } 
    \dateaccepted{ } 
    \datepublished{ } 
    
    \Title{TRACK: Telemetry-Based Racing Analysis and Coaching Kit in Sim Racing Games}
    
    \Author{Efe Çangırılı$^{1}$\orcidA{} and Murat Kurt $^{2*}$\orcidB{}}
    
    \AuthorNames{Efe Çangırılı and Murat Kurt}
    
    \address{%
    $^{1}$ \quad International Computer Institute, Ege University, \.{I}zmir, T{\"u}rkiye; efecangirili@gmail.com\\
    $^{2}$ \quad International Computer Institute, Ege University, \.{I}zmir, T{\"u}rkiye; murat.kurt@ege.edu.tr}
    
    \corres{Correspondence: murat.kurt@ege.edu.tr; Tel.: +90-232-311-3227 (M.K.)}
    
    \abstract{This paper presents TRACK (Telemetry-Based Racing Analysis and Coaching Kit), which is a framework for analyzing driving performance in sim racing and profiling how individual drivers behave behind the wheel. We report this framework together with its limitations: we calibrate each clustering result against a null, and when one does not separate from chance, we say so. Instead of restricting ourselves to scoring drivers or sorting them into preset labels, we represent each recording session as a compact geometry in a four-dimensional behavioral space (speed, braking, strategy, and consistency), and we group these fingerprints by their similarity using unsupervised clustering. Over time, we have developed and refined this framework on the open Assetto Corsa Gym (ACGym) dataset. This is a dataset that contains high-frequency, multi-channel telemetry from several drivers, cars, and tracks, and we additionally examine the framework on Assetto Corsa Competizione (ACC) sessions that the first author recorded through MoTeC. Our study suggests that corner types differ along a behavioral dimension that was not used to define them. It also suggests that when the car changes, only speed and consistency carry over in the restricted population, while repeatability could not be shown there for any of the braking or strategy measures. Cluster separation becomes less distinct as the range of available telemetry widens. Until that repeatability is shown, grouping on the braking and strategy dimensions cannot treat the car as interchangeable, which divides an already small sample into smaller cells. It is also not clear whether a driver's grouping carries over from one corner type to the next. We also normalize each metric against a reinforcement-learning reference agent. The reference does not depend on the sample, so the scale does not shift when the sample does. We intend these results as an analytical foundation for a personalized improvement suggestion system. The sample is small. The cross-car result changes when the sample is defined more broadly. These outcomes are preliminary.}
    
    
\begin{document}
    \section{Introduction}
    \label{intro}
    
    In motorsports, a driver's performance is most often reduced to a single number or a score: the lap time. Lap time is unambiguous and directly comparable, and that clarity is useful: a single number lets a driver see where they stand and gives them a concrete target to chase. Lap time does not give direction. It reports that one lap was faster or slower, but not how or why it was driven that way, nor what a driver might change to improve. Reduced to a lap time, or to a single overall skill score, drivers with very different styles are flattened into one ranking. That different styles can sit behind the same result is not hypothetical: analyzing open data from expert race drivers, Kegelman et al.~\citep{kegelman2017insights} found that two drivers in the same car could adopt clearly different styles, with different lines and speed profiles and one of them favoring a ``slow-in, fast-out'' approach, yet recorded comparable lap times. For ranking, this is no shortcoming. For improvement, the gap matters: a driver who knows only that they were slower still does not know what to work on in their own driving. This work asks how a driver actually drives, meaning their individual style, rather than only how fast, because that forms the basis for directed, personalized improvement.
    
    Modern racing simulators are, in effect, instruments built to capture driving in fine detail. Even a single session yields high-frequency, multi-channel telemetry (speed, throttle, brake, steering, and much more), sampled many times per second under controlled, repeatable conditions and with none of the cost, travel, or risk of injury and damage of real track testing. Read only at the surface, these channels show what the car was doing moment to moment; the harder question is what they mean. The goal is always to go faster, but the substance lies in how a driver got there: where and how they braked, the lines and trade-offs they chose, how steadily they repeated them. These elements set one driver apart from another. On its own, though, this profusion of data is closer to a bottomless well than to an answer. It is continuous and multi-dimensional, and it is tied to the particular car and circuit it was recorded on. No single channel says how one driver differs from another; the value is in how the channels are organized. From this raw behavior, we distill four dimensions: speed, braking, strategy, and consistency. Together they describe how a driver drives. Each is operationalized through the metrics defined in Section~\ref{sec:fingerprint}. We compute a behavioral fingerprint for each recording session: a point in the four-dimensional behavioral space defined by these dimensions. We then group these fingerprints by their similarity, letting structure emerge from the data rather than from skill labels fixed in advance. We call this framework TRACK (Telemetry-Based Racing Analysis and Coaching Kit).
    
    The drivers who stand to benefit from this are not only professionals. Casual and competitive video-game players, amateur and semi-professional racers, and anyone trying to improve on track can gain from analysis of this kind, because it lets a driver do more than chase a ranking or a spot: they can see where their own style sits among others and work on it deliberately. The value is concrete. An amateur or semi-professional racer rarely has time to learn every car and circuit before an event, and may carry the same mistakes and habits from one race into the next; analysis that makes those patterns visible is more readily available to well-resourced competitors, and a system built on it could, in time, turn such patterns into directed coaching. Because comparable telemetry channels are also recorded in real cars, an approach developed in simulation might eventually inform coaching beyond it, particularly at the junior and club levels. In practice, this kind of analysis is often tied to specialist expertise and proprietary workflows: inside well-resourced teams, or behind subscription services that do not reveal how they work. An open, documented, and automatic method points the other way: it can make this kind of insight broadly available.
    
    Two strands of previous work are most relevant here, and each leans on a label our question does not have. The first identifies who is driving, treating the telemetry as a signature unique to a person and training models to recognize it, rather than to describe how that person drives. The second works from outcomes: it groups laps or drivers by result, and the analysis asks what distinguishes faster from slower performance. The closest study to ours, Hojaji et al.~\citep{hojaji2024aienabled}, clusters laps by lap time into faster and slower groups and predicts lap time, then reads off the behaviors that mark the faster laps. That grounds style in a measurable outcome, but it still anchors the analysis to lap time rather than recovering style on its own. These strands have produced useful methods and findings, and we build on them rather than putting them aside. Unsupervised profiling of driving style already exists, but mainly in everyday and naturalistic driving (separating, say, aggressive from cautious drivers for safety or insurance) and in general driver-training simulators, rather than in competitive racing. What remains comparatively open is characterizing an individual's racing style as an unsupervised fingerprint, drawn from raw channels such as throttle, brake, and steering, and presupposing neither a lap time nor a skill label. The question that follows is how that style is structured: whether it holds or varies across corner types and across cars. We aim to describe how each driver drives, not to rank who is faster. That description is the footing on which style-specific guidance could later stand. We review these strands in detail in Section~\ref{sec:related}.
    
    We developed this approach on an open high-frequency telemetry dataset that spans three circuits and three car types, and we draw on our own recorded sessions to add a second platform. The contributions of this work are the following:
    \begin{itemize}
        \item a representation of each recording session as an unsupervised, multi-dimensional behavioral fingerprint, grouped by the natural clusters these fingerprints form rather than by predefined labels;
        \item per-corner-type profiling, which measures whether a driver's grouping carries from one corner type to another; the corner types are shown to separate along a behavioral dimension not used to define them, while the agreement between the within-type groupings does not differ from chance in the restricted population;
        \item an analysis of cross-car repeatability, which finds that repeatability is confined, in the restricted population, to the speed and consistency dimensions, with none of the braking or strategy metrics surviving correction there;
        \item a coverage ladder that measures how cluster separation behaves as the range of available telemetry widens, and finds the null-corrected separation declining at every step in both populations; and
        \item a normalization that expresses each driver's metrics relative to a reinforcement-learning reference agent, inverting the human-normalized-score idea~\citep{mnih2015humanlevel}, reducing dependence on the composition of the sample.
    \end{itemize}
    Two limitations frame these results. First, they rest on a limited number of distinct drivers, which limits how far they can be generalized; they should be read as preliminary rather than definitive. Second, the framework is intended as an analytical foundation for a personalized improvement suggestion system, not as that system itself; building and evaluating the suggestion component is left to future work (Section~\ref{sec:futurework}). The remainder of the paper is organized as follows: Section~\ref{sec:related} reviews related work, Section~\ref{sec:methods} describes the data and methods, Section~\ref{sec:results} presents the results, Section~\ref{sec:discussion} discusses their implications and limitations, Section~\ref{sec:conc} states what a larger study would need to measure first, and Section~\ref{sec:futurework} sets out future work.

    \section{Related Work}
    \label{sec:related}
    
    As autonomous driving and agent-based systems have grown in prominence, racing simulators have become a natural proving ground for them: a high-fidelity, low-risk environment for developing driving agents. Much recent work follows this line, training reinforcement-learning agents to control a car at, and beyond, expert lap times \citep{wurman2022outracing, vasco2024superhuman}. Successful as this work is, its goal is control of the car, not the driver behind the wheel. Reading a person through telemetry has instead been approached along two main strands: one asks who is driving, the other how well: what separates a faster drive from a slower one.
    
    \subsection{Driver Identification from Telemetry}
    \label{sec:rw-id}
    
    The first strand asks a simple question: who is driving? It reads telemetry as a personal signature. Early work on road cars showed that signals from the internal bus of a vehicle, the network its components use to communicate, can single out a driver from just a short stretch of ordinary driving \citep{enev2016automobile}. Deep models further sharpened this: convolutional and recurrent networks over the same signals, reaching very high identification rates \citep{hu2023driver}. What these approaches share is their purpose: recognition. They treat a driver's style as a fingerprint to be matched against an identity, rather than something to be understood and improved. We ask something different: not who drove a lap, but how it was driven.
    
    \subsection{Performance and Outcome Modeling}
    \label{sec:rw-outcome}
    
    In this second strand, the question is not who is driving but who drives well, and the analysis works backward from that outcome. It is also the strand that comes nearest to our own setting. Hojaji and colleagues come at this question across three studies. Two of them were run on a professional racing simulator, with 93 and 174 participants, and the third works from telemetry gathered on public servers. In all three the laps are clustered by lap time and the behaviors that mark the quicker groups are read off those clusters; in the largest, lap time is also predicted, at an accuracy of 97.2 percent \citep{hojaji2022modeling, hojaji2023aiapproach, hojaji2024aienabled}. Remonda et al. apply the same logic in a professional racing simulator, grouping laps into performance levels and asking which control features separate them \citep{remonda2021comparing}. It could be argued that driving style is already well covered here. These studies do look closely at behavior. What they see, though, is behavior conditioned on an outcome label; a slow group can hold very different ways of driving side by side, and the heart of the analysis sits with the result, not with the driver. Where these lines converge, and what we build on, is that sim racing telemetry carries a reliable performance signal. We start one step earlier: from the style itself, before any outcome label, class, or prior description is attached, so that improvement can be sought within a driver's own way of driving, not only toward a faster number.
    
    \subsection{Unsupervised Driving-Style Clustering}
    \label{sec:rw-style}
    
    Beyond these two strands, unsupervised clustering of driving style is itself well established, but it has grown up largely outside racing. In everyday road driving, styles have been shown to emerge from vehicle kinematics alone: Chen and Chen clustered naturalistic trips into three styles, separated by the speed maintained, by braking, and by lateral and longitudinal acceleration \citep{chen2019driving}, and a separate line of work clusters style and skill together, asking which features best support the grouping \citep{chen2023feature}. Driver-training simulators tell a similar story. A recent system scored 412 records retained from 538 driving sessions on a training simulator, clustered the scores into three profiles, and put those profiles forward as the basis for profile-specific feedback \citep{medarevic2024simulation}. Racing has expertise studies of its own, but there the groups come first: sim racers are divided into a higher- and a lower-skilled group by the top and bottom quartiles of their fastest laps, with the middle half set aside \citep{joyce2024less}, and racing drivers are compared against non-racing drivers \citep{vanleeuwen2017differences}, so the structure is fixed before the data are seen. Missing is the combination: an unsupervised style fingerprint in competitive racing, presupposing neither a lap time nor a skill label, and a view of how that style is organized across corner types and across cars. This paper sits in that gap.

    \subsection{Positioning}
    \label{sec:rw-position}

    \begin{table}[htbp]
    \footnotesize
    \setlength{\tabcolsep}{4pt}
    \renewcommand{\arraystretch}{1.15}
    \caption{The studies reviewed above, set side by side on the axes that separate them from the present work.\label{tab:positioning}}
    \begin{tabularx}{\textwidth}{@{}>{\raggedright\arraybackslash}p{2.9cm} >{\raggedright\arraybackslash}p{1.8cm} >{\raggedright\arraybackslash}X >{\raggedright\arraybackslash}p{1.8cm} >{\raggedright\arraybackslash}p{2.2cm} >{\raggedright\arraybackslash}p{1.6cm}@{}}
    \toprule
    \textbf{Study} & \textbf{Unit grouped} & \textbf{Grouping built from} & \textbf{Label needed first} & \textbf{Setting} & \textbf{Style structure examined}\\
    \midrule
    Enev et al.~\citep{enev2016automobile} & Driver & In-vehicle bus signals & Identity & Road & ---\\
    Hu et al.; Azadani and Boukerche~\citep{hu2023driver, azadani2021driver} & Driver & The same signals, deep models & Identity & Road & ---\\
    Hojaji et al.~\citep{hojaji2023aiapproach, hojaji2024aienabled} & Lap & Telemetry, 93 and 174 participants & Lap time & Professional racing simulator & ---\\
    Hojaji et al.~\citep{hojaji2022modeling} & Lap & Telemetry scraped from public servers & Lap time & Sim racing, public servers & ---\\
    Remonda et al.~\citep{remonda2021comparing} & Lap & Lap time & Performance level & Professional racing simulator & ---\\
    Chen and Chen~\citep{chen2019driving} & Trip & Vehicle kinematics & None & Naturalistic road & ---\\
    Chen et al.~\citep{chen2023feature} & Driver & Kinematics & Driver ID (proxy) & Naturalistic road & ---\\
    Medarevic et al.~\citep{medarevic2024simulation} & Driver & Simulator scores, 412 records & None & Training simulator & ---\\
    Joyce et al.~\citep{joyce2024less} & Driver & Fastest lap & Skill group & Sim racing & ---\\
    van Leeuwen et al.~\citep{vanleeuwen2017differences} & Driver & Group membership & Racing or non-racing & Simulator & ---\\
    \midrule
    This work & Recording session & 19 raw telemetry metrics & None & Competitive sim racing & Corner type, car\\
    \bottomrule
    \end{tabularx}
    \end{table}

    Table~\ref{tab:positioning} sets these studies side by side. Of the ten, eight require a label before the analysis even begins: an identity in the first strand, an outcome in the second, and a skill group in the racing studies of the third. The two that require none work outside competitive racing. In one of the eight the label is indirect: the grouping itself is unsupervised, and the questionnaire the study is named for is applied after it, but the features the grouping runs on are selected through a supervised model whose label is driver identity \citep{chen2023feature}.

    The closest study to ours is Hojaji and colleagues: the kind of telemetry and the simulator environment are the same, and driving behavior is read directly from the telemetry channels. The difference lies in the unit: their unit is the lap, and the grouping is built from lap time; therefore, what the analysis recovers is the behavior that accompanies a faster outcome. The team that published the ACGym dataset we analyze is also the team behind one of the studies we compare against.

    Our distinction lies here: instead of taking the lap as the baseline, we group the recording sessions as a whole, and we build the grouping from 19 metrics without applying any labels. Thus, corner type and car cease to be conditions fixed in advance and become questions that the data can answer.

    We should also note that the question we are trying to answer is not identical to that of the other studies; for this reason, claiming to be better would be unfounded and incorrect. If the outcome itself is what matters, grouping accordingly is the straightforward path; our approach becomes necessary only when the question is ``how they drive'' and not ``how fast.''

    \section{Materials and Methods}
    \label{sec:methods}
    
    \subsection{Datasets}
    \label{sec:data}
    
    Our primary data source is the Assetto Corsa Gym (ACGym) dataset \citep{remonda2024acgym}, an open collection of high-frequency driving telemetry recorded in the Assetto Corsa simulator. We chose this dataset for three main reasons: it is distributed under a CC BY 4.0 license, so the analyses built on it can be reproduced and extended; it covers many distinct human drivers recorded under controlled, repeatable conditions; and it spans several cars and circuits, exactly the variety a study of driving style requires. Beyond its human portion, the dataset also ships with telemetry from a reinforcement-learning agent trained on the same circuits. These agent recordings are kept out of the human clustering population; they later serve as the external reference for our normalization (Section~\ref{sec:rlnorm}).
    
    ACGym provides multi-channel telemetry sampled at 50~Hz, including, but not limited to, speed, gear, engine state, throttle, brake, steering, and track position. The human portion is documented as comprising 15 distinct drivers whose experience ranges from professional to beginner, driving on three circuits (Monza, Barcelona, and the Red Bull Ring) in three car types with markedly different handling characteristics: a GT3-class car (BMW Z4 GT3), an open-wheel formula car (Dallara F317), and a low-powered road car (Mazda Miata). The distribution also covers a fourth circuit, Indianapolis. A raw human session is present there, but it is not represented in our processed feature set, so the widest of the evaluation tiers defined in Section~\ref{sec:evaldesign} adds only agent coverage. Table~\ref{tab:datasets} summarizes the composition.
    
    Not every identifier in the human portion corresponds to a human driver. Three were removed before analysis: an in-game artificial driver, the output of a policy ensemble, and a classical-control baseline that the dataset paper describes as such. Only the last is named in the documentation; the first two are identifiable from their identifiers alone. The artificial driver and the control baseline were still present in the corner-level feature tables and were removed there, while the ensemble output had been filtered at an earlier stage. A further three identifiers share a naming pattern found nowhere else in the dataset, with cyclically permuted numeric suffixes, which raised doubt about whether they record human sessions. Their status could not be settled from the data. We therefore carry both readings throughout: a restricted population that omits them, treated as primary, and an extended population that retains them, reported alongside it. Laps failing basic validity checks, such as incomplete laps and malformed recordings, were removed from both sources.
    
    The driver identifier distributed with the telemetry names a recording session, dated to a single day, rather than a person; the same individual can appear under several identifiers. The account field embedded in the telemetry filenames does not resolve this, since it records the simulator account under which a session was collected. We recovered a person-level grouping by matching the trailing initials of the identifiers. The grouping is our inference; no label in the data confirms it. Analyses of corner types, tiers, and the agent reference are therefore carried out at the level of recording sessions. Only the cross-car analysis requires the person-level grouping, and its conclusions are conditional on it.
    
    The grouping returns thirteen groups when those three identifiers are omitted and sixteen when they are retained, against the fifteen human drivers the dataset documents. Their status is therefore bounded, even though it is not settled: at most two of the three can be drivers not already among the thirteen, and at least one must repeat a driver already counted or record something other than a human session. That bound is only as good as the grouping. Two drivers who share initials would be merged into one group; one driver recorded under two spellings would be split across two. Nothing in the data distinguishes either error from a correct grouping.
    
    To broaden the sample and examine the framework from a second platform, we additionally draw on sessions recorded by the first author in Assetto Corsa Competizione (ACC) through the MoTeC telemetry suite: 16 complete laps across Monza, the N\"urburgring, and Silverstone in six GT3-class cars. A seventeenth recording was a byte-identical copy of one of these and was excluded. These sessions play a deliberately light, supporting role. They are not part of the main clustering population, and they are used only to show that the pipeline runs end to end on telemetry recorded outside the corpus it was developed on (Section~\ref{sec:results}).
    
    \begin{table}[htbp]
    \footnotesize
    \setlength{\tabcolsep}{4pt}
    \renewcommand{\arraystretch}{1.15}
    \caption{Composition of the data sources used in this study. Session counts are raw, before the exclusions described in the text.\label{tab:datasets}}
    \begin{tabularx}{\textwidth}{@{}>{\raggedright\arraybackslash}p{2.4cm} X >{\raggedright\arraybackslash}p{3.2cm} >{\raggedright\arraybackslash}p{3.0cm} >{\raggedright\arraybackslash}p{1.1cm}@{}}
    \toprule
    \textbf{Source} & \textbf{Circuits} & \textbf{Cars} & \textbf{Recording sessions} & \textbf{Rate}\\
    \midrule
    ACGym (primary) & Monza, Barcelona, Red Bull Ring & BMW Z4 GT3, Dallara F317, Mazda Miata & 18 / 24 / 19 per circuit & 50~Hz\\
    RL reference & The same three circuits, and Indianapolis & Same & Single agent & 50~Hz\\
    Own sessions (secondary) & Monza, N\"urburgring, Silverstone & Six GT3-class cars & 1 driver, 16 laps & --\\
 \bottomrule
\end{tabularx}
\end{table}

    \subsection{Corner Segmentation}
    \label{sec:corners}
    
    All analyses operate at the level of individual corner passes rather than whole laps, because the behaviors that distinguish drivers, such as where braking begins, how much speed is carried through the apex, and how early the throttle reopens, live in corners.
    
\begin{figure}[!htb]
\centering
\includegraphics[width=\textwidth]{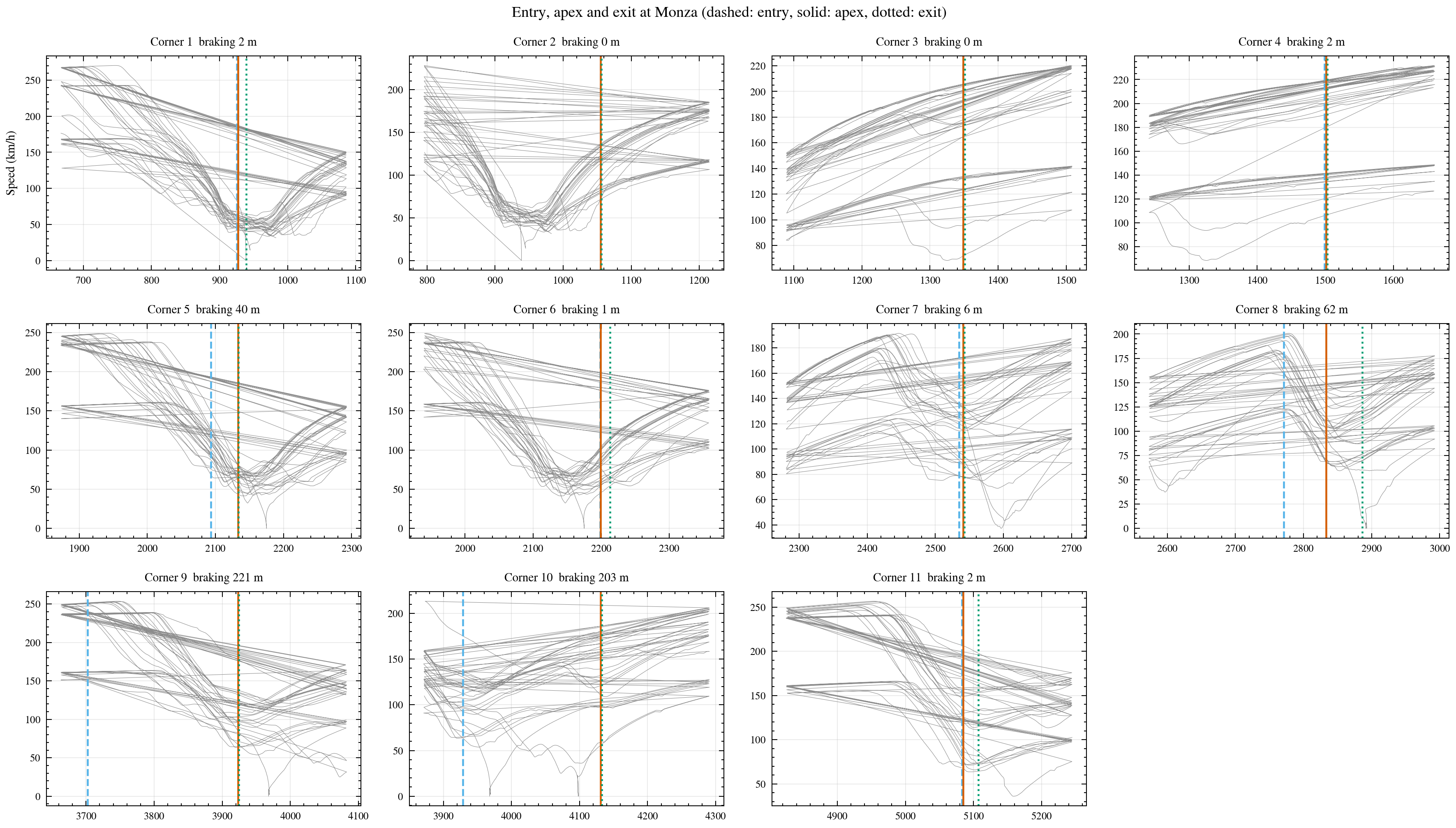}
\caption{Entry, apex, and exit points located at Monza for one recording session. Each panel shows one corner; the grey traces are the session's laps.\label{fig:segmentation}}
\end{figure}

    Corner candidates are located from a weighted combination of four signals, namely path curvature, deceleration, lateral acceleration, and brake application, computed along a single reference lap, with a candidate retained when at least two signals agree. A second pass recovers high-speed corners that produce curvature without appreciable braking, and a third, reference-guided pass completes the set to the published corner count of the circuit. The provenance of every corner is recorded, and the counts obtained at each stage are reported in Section~\ref{sec:results}. Across the three circuits, this yields 37 distinct corners, each with its own entry, apex, and exit reference points. Figure~\ref{fig:segmentation} shows the
result for one session at Monza. Each panel is one corner, with the session's laps drawn in grey behind
the three reference points, so the spread the boundaries have to accommodate can be read directly: the
points are fixed per corner, the lines through them are not.
    
    Each corner pass is segmented into five phases spanning approach to exit. Phase boundaries follow the driver's own inputs rather than fixed distances: braking onset opens the braking phase, and the exit point is located by a three-tier rule, namely the first sustained full-throttle application where present, a throttle-only criterion otherwise, and a distance-based fallback when neither is available. Segmenting by behavior rather than by geometry alone keeps the phases comparable across drivers with very different styles.
    
    Finally, corners are classified into slow, medium, and fast types by their characteristic speed, taken as the apex speed recorded for each corner in the reference corner table, using global thresholds at the 33rd and 66th percentiles of the pooled distribution and shared across all circuits. The resulting partition is insensitive to the exact quantile definition: using exact tertiles instead yields the same assignment of every corner. A single global scale, rather than per-track thresholds, is what makes corner types comparable across circuits: a medium corner at Monza and a medium corner at Barcelona mean the same thing.

\subsection{Behavioral Fingerprint}
\label{sec:fingerprint}

To characterize how a driver drives, we evaluate the behaviors they favor across different corner types along four dimensions, each capturing a complementary face of style. The first is speed: how fast a driver enters a corner, how much speed is carried through the apex, and how quickly the car leaves. The second is braking, which we read from where the brake goes on and how it comes off; the release matters as much as the application, since it shapes how the car turns in. Third comes strategy, the choices that shape a corner beyond raw pace: how long braking extends, and how much the car is allowed to coast before the throttle is reapplied. The fourth, consistency, asks a different question: does a driver take every corner of the circuit at a similar pace, or does that pace spread out from corner to corner? Speed and braking describe what is done in a single pass; strategy and consistency describe how those passes are distributed across the circuit.

\begin{figure}[!htb]
\centering
\includegraphics[width=0.88\textwidth]{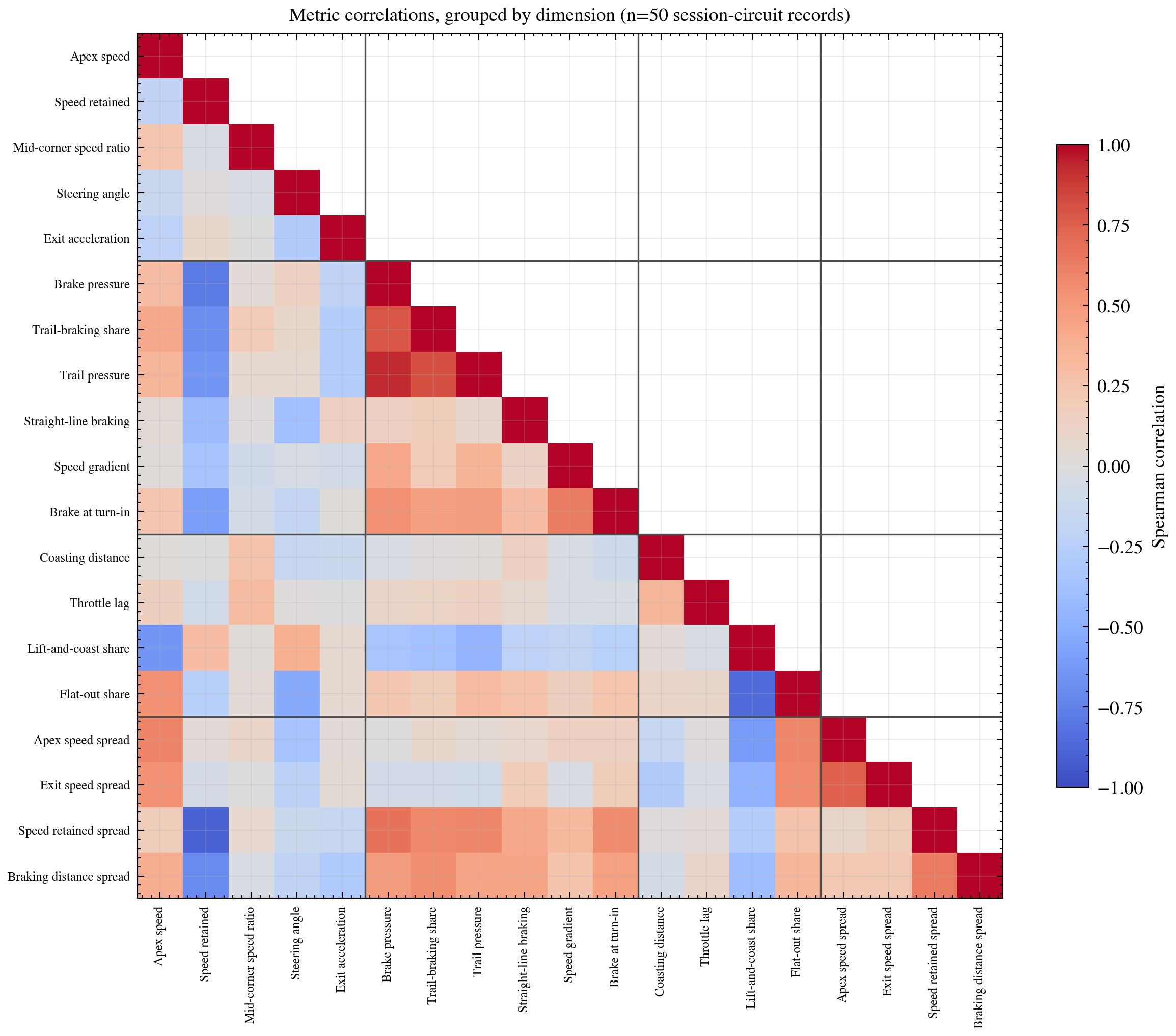}
\caption{Rank correlations between the 19 metrics, ordered by dimension, over the 50 sessions and circuit records of the restricted population.\label{fig:corr}}
\end{figure}

Each dimension is quantified by a set of metrics computed over corner passes and aggregated per recording session (Table~\ref{tab:metrics}). The fingerprint uses 19 metrics in total: five for speed, six for braking, four for strategy, and four for consistency. The four dimensions are defined in advance, and the metrics are placed under them. Candidate metrics carrying almost the same information were pruned by correlation before the set was fixed. Some correlation remains among those retained; these were kept deliberately, as they measure distinct aspects of the same behavior. The grouping itself was checked by comparing, for every metric, its correlations with the metrics of its own dimension against its correlations with the metrics outside it. Figure~\ref{fig:corr} orders the metrics by dimension, so
the check can be read as block structure rather than as a list of pairwise comparisons: the strategy and
braking blocks stand out against their surroundings, while the speed block does not separate from the
rest. The comparison is passed by 11 of the 19 metrics under both population definitions, and the eleven are not spread evenly over the four dimensions. In the restricted population, all four strategy metrics are more closely related to their own dimension than to the rest, and five of the six braking metrics are. The consistency dimension divides two against two. In the speed dimension, no metric passes at all: apex speed correlates at 0.24 inside its dimension and at 0.65 outside it, and the widest gap belongs to the speed retained at the apex, at 0.20 inside against 0.89 outside. That gap has an architectural cause, since the spread of the same quantity across corners is placed under consistency by design, and the two metrics are therefore computed from one measurement. The extended population moves one metric into the speed dimension and one out of strategy, and it leaves the total unchanged. Because the raw metrics live in their own units and on their own scales (one in km/h, one in meters, one a dimensionless ratio), what each carries on its own is limited, and together they do not form a common fingerprint. The metrics within a dimension are averaged into one dimension score, and that score is then scaled across the population so that the four dimensions share a common range. Because the averaging precedes the scaling, a dimension score is driven mainly by whichever of its metrics varies over the widest numerical range: straight-line braking distance in meters weighs more than brake pressure on a zero-to-one scale. Clustering thus operates on four dimension scores rather than on the raw metrics. This within-population scaling is a separate layer from the reinforcement-learning reference introduced in Section~\ref{sec:rlnorm}.

Corner-type profiling repeats the same logic on a reduced set of metrics. To see how a driver behaves in slow, medium, and fast corners separately, the fingerprint is recomputed within each corner type. A circuit, however, holds only a few corners of each type, and over so few passes some of the 19 metrics lose their reliability. Within a corner type, we therefore keep the metrics that remain meaningful even with a small number of corners: apex speed, braking distance, exit speed, and trail braking, each as a mean and, where applicable, a standard deviation. Across the three corner types, this gives 21 features per recording session. Because the computation requires only coverage of each corner type rather than presence on every circuit, it can draw on a larger set of recording sessions than the global fingerprint (Section~\ref{sec:results}).

We deliberately did not remove outlying records, and we distinguish this from data validity. Records that were degenerate, that is, non-human or malformed telemetry, were excluded, but a driver whose profile simply sits far from the others was kept. Such a profile can be read as noise, yet these are real drivers, and an unusual profile is as likely to be style as noise. A behavior seen in a single driver here may also correspond to an entire subpopulation in larger datasets; discarding it would mean setting aside not one driver but one kind of driver.

Each recording session is therefore represented as a point in a four-dimensional behavioral space: a compact geometry that stands in for the driver's style, and the object the clustering stage operates on.

\begin{table}[htbp]
\footnotesize
\setlength{\tabcolsep}{4pt}
\renewcommand{\arraystretch}{1.15}
\caption{The 19 metrics that make up the fingerprint, grouped by dimension. Spread metrics are computed across the corners of a circuit. The last column gives the scaling each metric receives in Section~\ref{sec:rlnorm}: a ratio to the agent reference, or scaling within the population.\label{tab:metrics}}
\begin{tabularx}{\textwidth}{@{}>{\raggedright\arraybackslash}p{4.3cm} >{\raggedright\arraybackslash}p{1.7cm} X >{\raggedright\arraybackslash}p{1.6cm}@{}}
\toprule
\textbf{Metric} & \textbf{Unit} & \textbf{Definition} & \textbf{Scaling}\\
\midrule
\multicolumn{4}{@{}l}{\textit{Speed}}\\
Apex speed & km/h & Speed at the apex reference point. & ratio\\
Speed retained at the apex & --- & Apex speed as a fraction of the speed at braking onset, so a higher value means less speed was given up. & ratio\\
Mid-corner speed ratio & --- & Lowest speed as a fraction of the highest over a window centered on the apex. & ratio\\
Steering angle through mid-corner & degrees & Mean absolute steering wheel angle over the same window. & ratio\\
Exit acceleration & km/h per m & Change in speed per unit distance from the apex to the exit point. & population\\
\midrule
\multicolumn{4}{@{}l}{\textit{Braking}}\\
Brake pressure & 0 to 1 & Mean brake input from braking onset to the apex. & ratio\\
Trail-braking share & proportion & Share of corners in which braking continues past turn-in. & ratio\\
Trail-braking pressure & 0 to 1 & Mean brake input over the final part of the braking phase. & ratio\\
Straight-line braking distance & m & Distance from braking onset to turn-in. & population\\
Speed gradient, straight-line braking & s$^{-1}$ & Change in speed per unit distance over that phase. & population\\
Brake at turn-in & 0 to 1 & Brake input at the turn-in point. & ratio\\
\midrule
\multicolumn{4}{@{}l}{\textit{Strategy}}\\
Coasting distance & m & Distance covered with neither brake nor throttle applied. & ratio\\
Throttle lag & m & Distance from the apex to the first sustained throttle application. & ratio\\
Lift-and-coast share & proportion & Share of corners the reference table labels as lift-and-coast. & population\\
Flat-out share & proportion & Share of corners the reference table labels as flat-out. & population\\
\midrule
\multicolumn{4}{@{}l}{\textit{Consistency}}\\
Apex speed, spread across corners & km/h & Standard deviation of apex speed. & population\\
Exit speed, spread across corners & km/h & Standard deviation of exit speed. & population\\
Speed retained at the apex, spread across corners & --- & Standard deviation of the retained fraction. & population\\
Braking distance, spread across corners & m & Standard deviation of the distance from braking onset to the apex. & population\\
\bottomrule
\end{tabularx}
\end{table}

\subsection{Clustering and Validation}
\label{sec:clustering}

It is fair to ask why we chose unsupervised clustering here. Our argument is that structure should be drawn out of the data rather than imposed on it. When grouping recording sessions, we therefore use no skill labels, no lap times, and no comparable prior assumption as input. We evaluate candidate numbers of clusters against three internal validity indices: the silhouette coefficient \citep{rousseeuw1987silhouettes}, and the Calinski--Harabasz \citep{calinski1974dendrite} and Davies--Bouldin \citep{davies1979cluster} indices.

\begin{figure}[!htb]
\centering
\includegraphics[width=0.85\textwidth]{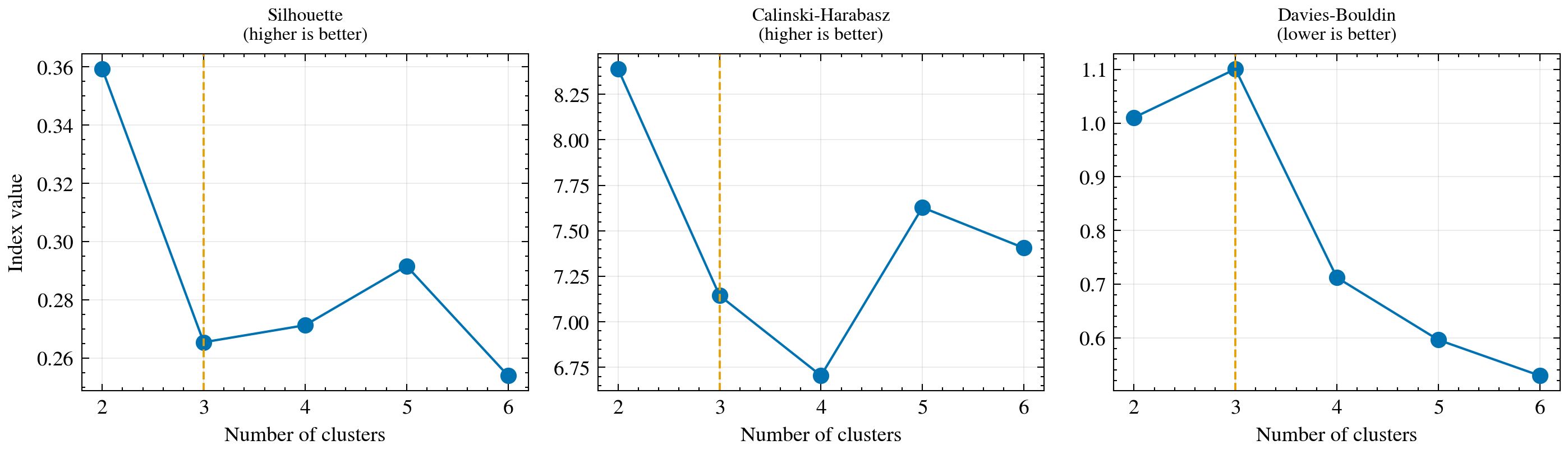}
\caption{Internal validity indices across candidate numbers of clusters. The indices do not agree on a single value.\label{fig:validity}}
\end{figure}

The indices do not converge on a single value: silhouette and Calinski--Harabasz favor the coarsest split, while Davies--Bouldin favors a much finer one, so the data do not single out one solution. Figure~\ref{fig:validity} plots the three indices across the
whole candidate range, which shows that the disagreement is not local to one value: the number we fix
sits at none of the three optima. We therefore fix the number of clusters at three for all analyses, so that partitions remain comparable across tiers and corner types, and report this as a design choice rather than as an index-selected optimum. At this sample size, the finest solution would leave roughly two records per group, too sparse to interpret, while the coarsest yields little more than a binary split. Sensitivity to this choice is reported in Section~\ref{sec:results}, together with the index values for every candidate.

All partitions were computed with a fixed random seed and 20 initializations; a 60-seed sensitivity analysis gives a standard deviation no greater than 0.014 in the silhouette coefficient and does not change the ordering reported in Section~\ref{sec:results}.

Because internal validity indices are sensitive to sample size, a silhouette value cannot be read on its own. For each configuration, we therefore construct a null distribution by independently permuting each feature column, which destroys the joint structure while preserving the marginal distribution of every metric, and recompute the coefficient under the same procedure over 1000 permutations. We report the observed value together with its excess over the null mean and the corresponding permutation $p$-value, computed as $(k+1)/(B+1)$, where $k$ counts the permutations reaching or exceeding the observed value and $B$ is the number of permutations. With $B = 1000$, the smallest attainable value is 0.001, and a $p$-value reported at that level should be read as a floor rather than as a point estimate. Variability is quantified separately. We draw 1000 subsamples, each containing 80\% of the records, without replacement; we avoid resampling with replacement because it inflates cluster validity indices. Each subsample yields its own coefficient value, and we report the 5th and 95th percentiles of those values. We call this a subsample range, not a confidence interval: the scheme does not support a coverage claim.

One further argument shaped the design: a single algorithm can impose structure of its own, so no single algorithm's output should be trusted on its own. We therefore work with three methods resting on different assumptions: a centroid-based partition, Ward's agglomerative hierarchy, and a fuzzy clustering that assigns graded memberships rather than hard labels, so that a record may belong partly to more than one group. Where these algorithms converge on the same partition, the partition is less dependent on the choice of method. Convergence alone does not make the structure a property of the data. Building the pipeline also gave us the chance to try other families of algorithms: a density-based method proved highly sensitive to its neighborhood parameter, labeling every record as noise at tight settings and recovering only two groups at looser ones while leaving several records unassigned.

Agreement between partitions is quantified with the adjusted Rand index \citep{hubert1985comparing}. The same index also lets us compare the clusterings obtained in different corner types. Both of the procedures just described are applied to it as well: the permutation null, and the subsample range. We further report a leave-one-out cross-validation score, which we treat as a measure of how separable the clusters are; it is not a claim of predictive performance, and the number of records would not support one. Because one person can contribute several recording sessions, we repeat the procedure with folds grouped by person, so that all sessions attributed to the same individual are held out together. Finally, to interpret what distinguishes the groups, we fit a supervised model to the cluster labels and read its feature attributions with SHAP \citep{lundberg2017unified}, which surfaces the metrics that set each group apart.

\subsection{Reinforcement-Learning Reference Normalization}
\label{sec:rlnorm}

The scaling described so far is relative to the population: a driver's metrics carry meaning only against the other drivers in the sample. With a small sample, this makes every value dependent on who else happens to be in it. The dataset also includes telemetry from a reinforcement-learning agent trained to optimize reward on the same circuits. Its metrics are computed in the same metric space as the human drivers, through the same segmentation and the same definitions; human and agent data are never mixed, and the agent is used only as an external reference. The reference is assembled per circuit from the agent sessions distributed with the dataset and segmented by the procedure of Section~\ref{sec:corners}.

The idea behind this is a reversal. Common practice in reinforcement learning is to use human-normalized scores, expressing an agent's performance as a fraction of a human baseline \citep{mnih2015humanlevel}; here the direction is inverted, and each human metric is expressed as a fraction of the agent's. A value of one means parity with the reference, and a value below one means the driver's value is the smaller of the two. The ratio records relative magnitude, not performance. The scale stays the same across drivers, circuits, and cars, because the denominator does not depend on the sample.

Not every metric admits a ratio. In three cases the division is either undefined or meaningless: where the reference value is effectively zero, where the quantity can take negative values, and where the metric measures dispersion or a bounded proportion rather than a level. These fall back to the population scaling described earlier. Of the 19 metrics, 10 are expressed as ratios to the reference and 9 keep population scaling. The consistency dimension consists entirely of dispersion metrics, so it carries no ratio component and remains population-relative by design.

\subsection{Evaluation Design}
\label{sec:evaldesign}

One question remains as this section closes. Do the clusters reflect structure in the data, or do they depend on the narrow slice the fingerprint was developed on?

To answer this we ran the pipeline over a nested series of tiers. The first covers a single GT3-class car across the three circuits. The second adds the open-wheel formula car, the third adds the low-powered road car, and the fourth widens the set with a further circuit. Each tier is a superset of the one before, and the pipeline itself is left unchanged: the same segmentation, the same metrics, the same clustering and validation run on every tier, with only the input set differing.

Across tiers we compare three things: how well separated the clusters are, how separable they remain under cross-validation, and whether the metrics that distinguish the groups stay the same. The fourth tier adds a circuit for which our processed feature set contains no human sessions, so in the human analysis it coincides with the third; we report it as such rather than presenting it as an additional level of evidence. A further consequence of this design is that the pipeline is not specific to the drivers or cars used here: it applies to telemetry recorded in the same form.

\section{Results}
\label{sec:results}

\subsection{Analyzed Population and Corner Structure}
\label{sec:respop}

The segmentation stage identified 37 corners across the three circuits: 16 at Barcelona, 11 at Monza, and 10 at the Red Bull Ring. Corner definitions are fixed per circuit and are derived independently of the analyzed driver set, so they are common to both population definitions used below. The set of corners entering a driver's measurements is narrower: the confidence filter removes between zero and three corners per driver, leaving 13 to 16 valid corners at Barcelona, 9 to 11 at Monza, and 9 to 10 at the Red Bull Ring. These ranges are identical under both population definitions. Of the 37 corners, 31 were located from the detection signals alone; the remaining six, four of them at Barcelona, were added in a reference-guided pass that completes the detected set to the published corner count of each circuit. Corner counts therefore match the published layouts by construction and are not offered as independent validation of the detector. The Barcelona figure corresponds to the pre-2023 configuration.

Corner type was assigned by partitioning the pooled apex-speed distribution at its 33rd and 66th percentiles (Section~\ref{sec:corners}). Across the 37 corners the apex speed used for typing averaged 119.07~km/h (SD 34.94, range 68.34--200.95), and the two boundaries fell at 106.77 and 127.31~km/h, giving 12 slow, 12 medium, and 13 fast corners. These groups are distributed unevenly over the circuits: 7 of the 12 slow corners lie at Barcelona, and 5 of the 13 fast corners at Monza (Table~\ref{tab:corners}). Five of the six reference-guided corners fall in the slow group. Every corner located from the detection signals carries at least two signal votes, while each of these six carries one, and in every case the same one, curvature; five of the twelve slow corners therefore rest on corners admitted below the threshold applied elsewhere, a composition we return to in Section~\ref{sec:restypes}.

Thirteen recording sessions in the restricted population are present at all three circuits; separately, thirteen inferred drivers appear in that population. The two counts coincide numerically but refer to different units, and the analyses below state which one they use. Those thirteen drivers produced twenty-one recording sessions in total: eight contributed a single session, and the remaining five account for the rest, so the recording sessions are not independent of one another within the population.

After the exclusions described in Section~\ref{sec:data}, the extended population retains 17 recording sessions at Monza, 23 at Barcelona, and 19 at the Red Bull Ring, against raw totals of 18, 24, and 19; the restricted population retains 14, 20, and 16. Sixteen sessions in the extended population are present at all three circuits, against the thirteen reported above for the restricted one.

The restricted population comprises 569 laps and the extended one 705. Coverage by car is uneven: the lowest-volume car contributes 65 of the 569 restricted laps and 65 of the 705 extended ones, and appears at Monza in only two sessions, a sparsity we return to in Section~\ref{sec:restiers}.

\begin{table}[htbp]
\footnotesize
\setlength{\tabcolsep}{4pt}
\renewcommand{\arraystretch}{1.15}
\caption{Corner inventory and type composition. Types are assigned from the pooled apex-speed distribution using global thresholds at 106.77 and 127.31~km/h. The standard deviation is computed with one degree of freedom removed.\label{tab:corners}}
\begin{tabularx}{\textwidth}{@{}X ccccc@{}}
\toprule
\textbf{Circuit} & \textbf{Corners} & \textbf{Slow} & \textbf{Medium} & \textbf{Fast} & \textbf{Reference-guided}\\
\midrule
Barcelona & 16 & 7 & 5 & 4 & 4\\
Monza & 11 & 2 & 4 & 5 & 1\\
Red Bull Ring & 10 & 3 & 3 & 4 & 1\\
\midrule
All & 37 & 12 & 12 & 13 & 6\\
\bottomrule
\end{tabularx}
\vspace{2pt}

\footnotesize Apex speed used for typing: mean 119.07~km/h, SD 34.94, range 68.34--200.95.
\end{table}

\subsection{Corner-Type Profiles}
\label{sec:restypes}

Thirty-seven corners were detected across the three circuits: sixteen at Barcelona, eleven at Monza, and ten at the Red Bull Ring. The corners were divided into three types by their apex speed, and the thresholds were taken as the 33rd and 66th percentiles of the corner population, at 106.8 and 127.3 kilometers per hour respectively. The division yields twelve slow, twelve medium, and thirteen fast corners.

The thresholds are computed from the apex speeds in the corner table, and not from the driver pool. As a result, the corner inventory and the type assignment are identical under both population definitions, and the only thing that changes between them is the set of drivers measured at those corners. The definition of a type is therefore held fixed in the comparisons that follow, and any difference observed there can be attributed to the driver set alone.

The corner types differ in behavior, and the trail-braking share measured by type shows it. As Table~\ref{tab:trail} shows, the share rises from slow to fast corners under both population definitions, and the medians preserve the same direction. In the means the two definitions agree closely at every corner type, with the largest gap, 0.0073, at fast corners. The medians separate, and they separate most at slow corners, where the gap of 0.0159 exceeds the 0.0139 observed at fast corners. The separation between the definitions is not particular to fast corners; it lies in the distance between the mean and median.

\begin{table}[H]
\setlength{\tabcolsep}{4pt}
\renewcommand{\arraystretch}{1.15}
\caption{Trail-braking share by corner type, averaged over recording sessions. Values are computed from the corner-type profile table; the two population definitions share the same corner inventory and differ only in the set of sessions.\label{tab:trail}}
\begin{tabularx}{\textwidth}{@{}X cccc@{}}
\toprule
\textbf{Corner type} & \textbf{Restricted mean} & \textbf{Restricted median} & \textbf{Extended mean} & \textbf{Extended median}\\
\midrule
Slow & 0.1421 & 0.1429 & 0.1415 & 0.1270\\
Medium & 0.1972 & 0.2000 & 0.1971 & 0.1972\\
Fast & 0.3197 & 0.2778 & 0.3124 & 0.2639\\
\bottomrule
\end{tabularx}
\vspace{2pt}

\noindent\footnotesize{Restricted population: 21 recording sessions. Extended population: 24.}
\end{table}

This gradient is consistent with the type division separating the corners along a behavioral dimension that was not used to define them. The comparisons that follow take this separation as given.

The drivers were clustered in two separate ways. In the pooled solution, each driver entered a single clustering with the twenty-one features measured across the three corner types. In the within-type solutions, each corner type was clustered separately with its own seven features. The number of clusters was set to three in both cases.

The pooled solution achieves a silhouette value of 0.188 in the restricted population and 0.211 in the extended population. The clusters contain six, twelve, and three records under the restricted definition, and eleven, ten, and three records under the extended definition. The three-member cluster consists of the same three records in both cases, and none of these records is among the three discussed below. The silhouette values of the within-type solutions lie between 0.229 and 0.276 in the restricted population and between 0.249 and 0.296 in the extended one.

The internal validity measures of these two solutions are not compared with one another; they are computed over different numbers of dimensions, and none of them has been calibrated against a null distribution at this sample size. They are given here as descriptive.

Whether a driver's profile carries across corner types is measured from the agreement between the within-type clusterings. Each corner type produces its own clustering, and the clusterings of two types are compared with the adjusted Rand index over the drivers measured in both. Three types give three pairs, and the value reported is the mean of those three.

In the restricted population, the mean agreement is 0.043, with a subsample range of $-0.011$ to 0.247 and a permutation value of 0.119. The agreement between the within-type clusterings therefore does not differ from chance. In the extended population, the same quantity is 0.271, with a subsample range of 0.056 to 0.346 and a permutation value of 0.001, which is the floor of the test rather than a point estimate. The primary reading of this comparison is a null result. The sensitivity reading is positive; however, this outcome holds only if the three records originate from three distinct drivers, a condition the design does not establish. The two ranges overlap broadly, so the difference between the two populations cannot be read from them alone. Both the ranges and the permutation values belong to the mean, and the individual pairs were not tested separately.

Within the restricted population, the two pairs involving slow corners are $-0.006$ and $-0.040$, both below chance. The only positive agreement, 0.174, falls between medium and fast corners. Within the extended population, the agreement between slow and fast corners rises to 0.484, while the other two pairs are 0.162 and 0.167. Adjacent types would be expected to agree more closely than distant ones, and that expectation is reversed only in the extended population, where the highest agreement of all falls between the slowest and the fastest corners.

The higher agreement in the extended population rests on three records. Their effect can be separated from the effect of any three. When three records drawn at random are removed from the extended population, the mean agreement falls from 0.271 to 0.210, with a standard deviation of 0.055 over three hundred repetitions and fifth and ninety-fifth percentiles of 0.109 and 0.294. When the three records that the restricted population omits are removed instead, the value falls to 0.043, which is lower than 299 of the 300 random removals and sits three standard deviations below their mean.

The effect does not reduce to a single record. Removing the single most influential record on its own brings the mean agreement to 0.160, and the drop produced by the three together is 2.06 times as large. Leave-one-out measurement understates the effect of the group: the mean individual effect of the three records is 0.033, against a mean of 0.037 across the population.

The drop falls unevenly across the pairs. In the two pairs that involve slow corners the agreement falls from 0.162 to $-0.006$ and from 0.484 to $-0.040$. In the pair of medium and fast corners it moves from 0.167 to 0.174, that is, it does not fall at all. This localization is descriptive, since the pairs were not tested individually.

Two explanations were examined and set aside. The three records are not outliers: their distances from the population center rank fourth, thirteenth and seventeenth among twenty-four, and their mean distance stands at 1.04 times the population mean. Nor does the effect come from individual atypicality, as the leave-one-out figures above show. The remaining explanation is that the three resemble one another and cluster together within each corner type. That reading was tested and is partly supported: the three fall in the same cluster at slow corners and fast corners, but not at medium ones. The pair formed by slow and fast corners is the one in which the three are together in both types, and it is also the pair whose agreement changes the most. The remaining two pairs each contain only one such type, yet they behave differently from each other. Co-membership therefore does not, on its own, account for the pattern reported above.

The grouping of records into persons is inferred rather than explicitly given. Consequently, the three records may represent repeated sessions by a single driver. In that case the observed agreement would reflect driver identity rather than the transfer of a profile across corner types. This second reading has not been tested here, and it is recorded as a hypothesis.

The results in this section should be read with three limitations in mind.

Corner detection works from a single reference lap and rests on a weighted combination of four signals, with at least two signals required to agree before a corner is accepted. Six of the thirty-seven corners did not meet that criterion and were completed by reference to the published layout, and each of the six carries a single signal vote. Five of these six fall into the slow category. The weakest signal support in the inventory is therefore concentrated there, in five of the twelve slow corners. The pairs on which the three records had the largest effect are exactly the pairs that involve slow corners. The link between the two facts has not been tested and is recorded here as a possible confound.

The clusterings treat each recording session as an independent observation. Section~\ref{sec:respop} has already shown that they are not. Working at the session level is a design decision declared in Section~\ref{sec:data}. We did not test whether sessions attributed to one driver fall into the same corner-type cluster. The agreement measured between corner types may therefore carry a component of driver identity that the design does not separate. The pooled solution shows that this can happen: two of the three records in the three-member cluster are sessions attributed to one driver, while that driver's third session falls in a separate cluster. As reported in Section~\ref{sec:restiers}, the check found no such leakage in the tier ladder. This outcome is specific to a separate clustering and does not apply to the corner-type clusterings reported here. The concern applies to the extended population most directly, since the three records on which the effect rests are also the records whose grouping is least certain.

The type boundaries are derived from the percentiles of this corner population, and the number of clusters was set to three rather than selected. A sample drawn from other circuits could change both choices.

\subsection{Behavior of the Fingerprint as Coverage Widens}
\label{sec:restiers}

The tiers are distinguished by what they contain rather than by how they are processed. The narrowest contains a single GT3-class car, with 8 recording sessions in the restricted population and 10 in the extended one. Adding the open-wheel car raises these to 13 and 16. Adding the low-powered road car does not add any further sessions, so that step changes only the cars in the set. The widest tier coincides with it in the human analysis, for the reason given in Section~\ref{sec:evaldesign}.

\begin{figure}[!htb]
\centering
\includegraphics[width=\textwidth]{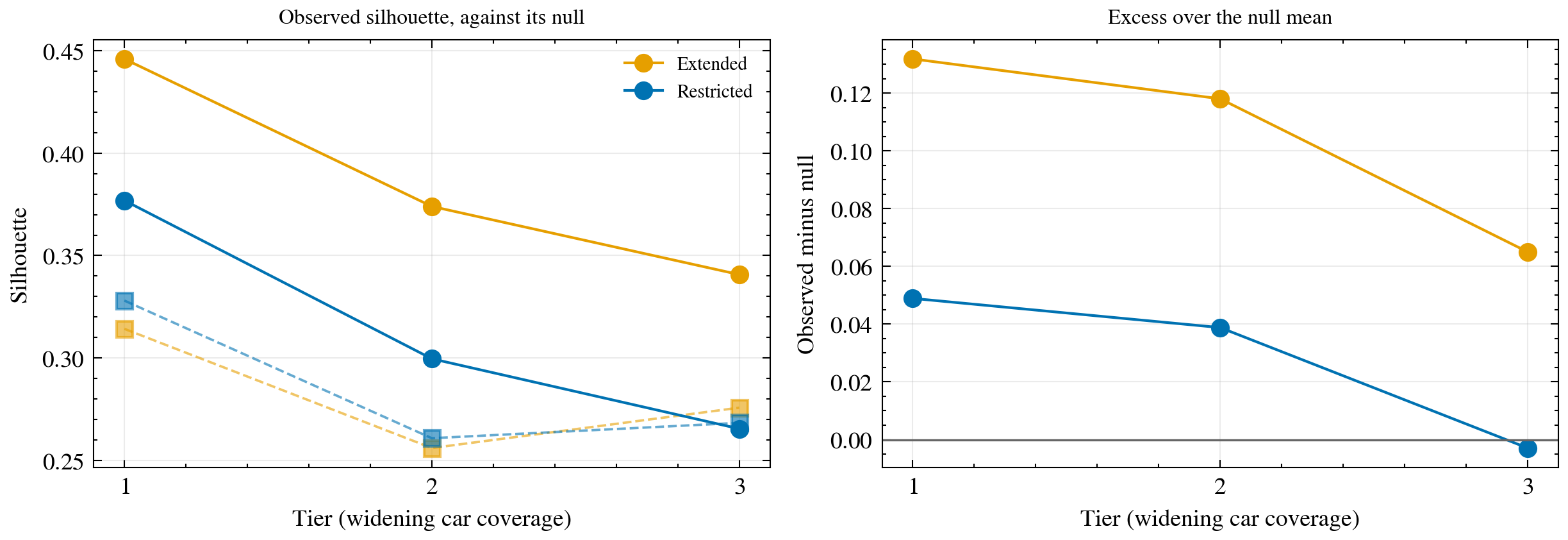}
\caption{Cluster separation across tiers. Left: the observed silhouette against its permutation null, shown as a dashed line. Right: the excess over the null mean.\label{fig:ladder}}
\end{figure}

The silhouette coefficient is not comparable across tiers by itself, since the permutation null rises as the sample shrinks: at the narrowest tier the null means are 0.328 and 0.314, against a range of 0.256 to 0.276 at the wider ones. A considerable portion of the apparent decline from the narrowest tier to the widest therefore reflects sample size rather than lost structure: 54 percent in the restricted population and 37 percent in the extended one. Each coefficient is reported against its own null (Table~\ref{tab:tiers}). Figure~\ref{fig:ladder} lets us read the observed coefficient and the null-corrected excess separately. In the left panel, the observed coefficient is drawn together with its permutation null, so that the movement of the null itself can be seen beside the apparent decline. The right panel drops the coefficient and plots the excess alone. That excess is the only quantity on the ladder that can be compared across tiers.

The comparison that can be made cleanly here is the one between the second and third tiers: both are run over the same records, and the only difference between them is the added car. The separation declines there on three different measures. In the restricted population, the coefficient falls from 0.300 to 0.265, the permutation $p$-value rises from 0.19 to 0.51, and the excess over the null even changes sign, from $+0.039$ to $-0.003$. In the extended population, the corresponding values are 0.374 to 0.341, 0.002 to 0.055, and $+0.118$ to $+0.065$. The null-corrected excess declines at every step of the ladder, in both populations.

These two populations differ in how much structure was present to begin with. In the restricted population, no tier separates from its null: the largest excess is $+0.049$ at the narrowest tier ($p = 0.22$), and at the widest the coefficient of 0.265 sits just below its null mean of 0.268. In the extended population, the two narrower tiers do separate ($p = 0.023$ and $p = 0.002$), while the widest is borderline, with an observed 0.341 against a null 95th percentile of 0.342.

In parallel, a second measure moves the other way over the same step. Leave-one-out accuracy rises from 0.62 to 0.69 in the restricted population and from 0.75 to 0.88 in the extended one when the road car enters. Over the full ladder it is not monotonic, since it dips at the second tier in both populations. Additional evidence addresses the leakage from sibling records that we had allowed for. When the folds are grouped by person, cross-validation yields identical accuracy across all eight combinations of tier and population, indicating that the leakage does not arise. The two measures answer different questions, one about how compact the groups are and one about how separable they are, and we report both without reconciling them.

Setting the number of clusters to three is a design decision, and it is important. In the case of two clusters, the opposite happens in the extended population, where the trend is reversed, going up from 0.329 to 0.399 along the ladder; two clusters is also the value that is preferred by the silhouette in seven out of the eight configurations. Between all three clustering methods, the difference in the coefficients is never higher than 0.028, although the agglomerative algorithm in the extended population is non-monotonic. Across 60 different random seeds, the standard deviation is never higher than 0.014, even though the seed we fixed creates the highest possible difference between the second and third tiers in the extended population and a below-average one in the restricted population.

The ladder therefore measures the behavior of an internal validity index as coverage increases. It does not prove that the fingerprint generalizes.

\begin{table}[htbp]
\footnotesize
\setlength{\tabcolsep}{4pt}
\renewcommand{\arraystretch}{1.15}
\caption{Cluster separation across tiers, each coefficient reported against its own permutation null over 1000 permutations. The widest tier is identical to the third in the human analysis and is omitted.\label{tab:tiers}}
\begin{tabularx}{\textwidth}{@{}X c c c c c c c@{}}
\toprule
\textbf{Population} & \textbf{Tier} & \textbf{$n$} & \textbf{Silhouette} & \textbf{Null mean} & \textbf{Excess} & \textbf{$p$} & \textbf{LOO acc.}\\
\midrule
Restricted & 1 & 8 & 0.3769 & 0.3280 & $+0.0489$ & 0.215 & 0.625\\
Restricted & 2 & 13 & 0.2996 & 0.2608 & $+0.0388$ & 0.191 & 0.615\\
Restricted & 3 & 13 & 0.2654 & 0.2683 & $-0.0029$ & 0.507 & 0.692\\
\midrule
Extended & 1 & 10 & 0.4460 & 0.3142 & $+0.1318$ & 0.023 & 0.800\\
Extended & 2 & 16 & 0.3740 & 0.2560 & $+0.1180$ & 0.002 & 0.750\\
Extended & 3 & 16 & 0.3407 & 0.2757 & $+0.0651$ & 0.055 & 0.875\\
\bottomrule
\end{tabularx}
\end{table}

\subsection{What Distinguishes the Groups}
\label{sec:resattr}

The metrics behind the grouping were read from a random forest fitted to the
cluster labels over the same 19 metrics, at the widest tier. The forest is a
surrogate, and its purpose is to make the partition readable. Its leave-one-out
accuracies are the ones already given in Table~\ref{tab:tiers}, and the interval
that belongs with a proportion of this size is wide: 9 of 13 records give a 95\%
Wilson interval \citep{wilson1927probable} of 0.424 to 0.873 in the restricted
population, and 14 of 16 give
0.640 to 0.965 in the extended one. The lower bound of each clears its
majority-class baseline of 0.385 and 0.375. Shuffling the cluster labels 200
times brings mean accuracy down to 0.297 (standard deviation 0.159) and 0.292
(0.140), which places the observed values at a label-permutation $p$ of 0.020 in
the restricted population and, in the extended one, at the smallest value 200
draws can produce, 0.005. This test is separate from the silhouette permutation
reported in Table~\ref{tab:tiers}: it asks whether the labels are learnable from
the metrics at all. The surrogate recovers the partition well beyond what
shuffled labels achieve, so the attributions below are read from a model that has
learned the grouping it explains.

Apex speed leads the attribution ranking in both populations. Refitting the
forest at 25 seeds leaves that position unchanged in every case, and the full
ranking tracks the reported one throughout: the rank correlation never falls
below 0.898. Read group by group, apex speed leads in four of the six groups, comes second
in one, and third in one. That last group is also the only one of the six in
which the two ways of computing the attribution disagree on the leader:
computed over the full sample it is the speed gradient in straight-line
braking, and computed within the group alone it is apex speed. A metric that
leads almost everywhere is the axis the partition is built along, and the
separation between groups is carried by what sits behind it. Those two positions are held by the same pair in both
populations, the speed gradient in straight-line braking and the spread of exit
speed across corners, with rank ranges of 2 to 3 and 2 to 4; which of the two
comes second is undetermined at this sample size. Below third place no metric
holds its position: throttle lag moves between ranks 4 and 9, straight-line
braking distance between 4 and 11.

\begin{figure}[!htb]
\centering
\includegraphics[width=\textwidth]{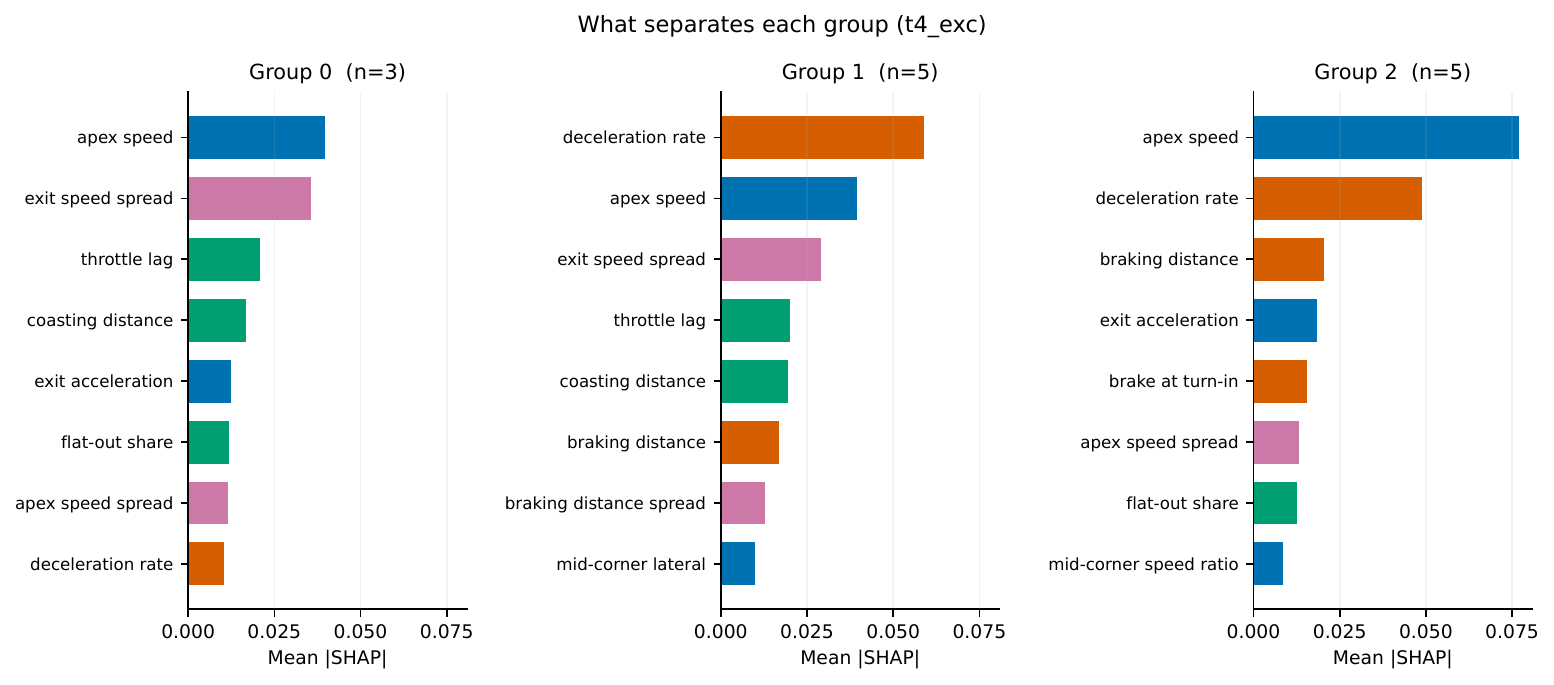}
\caption{Mean absolute SHAP value per metric, computed separately for each
group, in the restricted population. The eight highest metrics are shown per
panel and colored by dimension. The three-member group is the leftmost panel and
its ordering rests on three records. The corresponding panels for the extended
population are given in the Supplementary Material.\label{fig:attr}}
\end{figure}

Figure~\ref{fig:attr} sets the three groups of the restricted population side
by side. Read across the panels, the metric in second place is a different one
in each, and each belongs to a different dimension: the spread of exit speed
across corners, apex speed, and the speed gradient in straight-line braking.
The pooled ranking reports the leader and averages that difference away.

The leave-one-out scheme leaves sibling records in training. The 13 records
analyzed here are the sessions of the restricted population present at all three
circuits; they come from 10 inferred drivers, fewer than the 13 the population
holds in total, because one driver contributes three of these sessions and
another two. The person-grouped folds reported above address exactly this, and
the attribution inherits that result. The agreement between the
forest's own importance ranking and the attribution ranking ($\rho = 0.9789$ and
$0.9842$) measures the consistency of two readings of the same model, and carries
no information about driving. One group is led by the speed gradient in
straight-line braking. That position survives in only 15 of 25 seeds, which is
why no distinguishing metric is claimed for that group.

\subsection{Transfer Across Cars}
\label{sec:rescars}

This subsection sets out to answer a sole question: whether the same person can be recognized in a different car. The unit of analysis here is the driver rather than the recording session, and this is the only place in the paper where that grouping is required. The restricted population provides 90 car and circuit records from 13 inferred drivers, and the extended population provides 107 records from 16. One step in the design could not be measured. Matching an individual to their own records in the same car on a different day would separate the driver variable from the session variable, but the largest such cell holds only four individuals, too few to distinguish recognition from chance.

\begin{figure}[!htb]
\centering
\includegraphics[width=\textwidth]{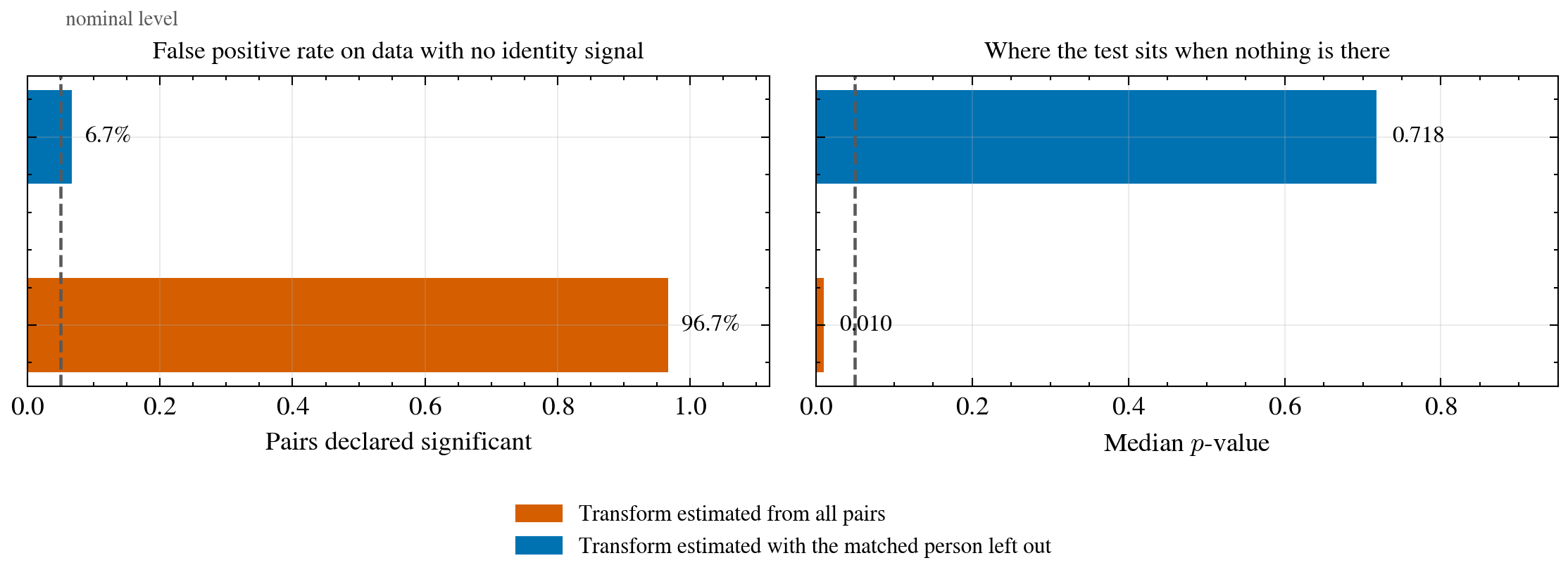}
\caption{Behavior of the two whitening procedures on synthetic data containing no identity signal, at 12 people, 19 metrics and 30 replications.\label{fig:calib}}
\end{figure}

In designing this measurement architecture, our experiments showed that a whitening transform estimated from the same pairs it is later used to match inflates the result. On synthetic data containing no identity signal, at the size of the largest car pairing available here (12 people, 19 metrics, 30 replications), the naive version declares 96.7 percent of pairs significant and returns a median $p$-value of 0.0099. Estimating the transform with the matched person left out returns a false positive rate of 6.7 percent and a median $p$-value of 0.718. Figure~\ref{fig:calib} compares the two procedures on that synthetic data. Since the data carry no identity signal, the correct outcome is the uninformative one, and only the corrected procedure returns it. The effect is also measurable in the observed data: under the naive transform five of the seven within-circuit comparisons reach significance, and under the corrected transform one does. Every figure reported below uses the corrected procedure.

In the restricted population, none of the three car pairings survives correction for multiple comparisons: the adjusted values are 0.191, 0.144, and 0.191. In the extended population, two of the three do survive, at 0.022 and 0.046, while the third does not, at 0.074 (Table~\ref{tab:crosscar}). The primary reading of this analysis is therefore a null result, and the sensitivity reading is a positive one.

The three pairings also disagree among themselves, and the disagreement follows the number of people available. The pairing with the most drivers, the GT3 car against the open-wheel car, returns the lowest accuracy in both populations: 0.25 against a chance level of 0.08, and 0.33 against 0.07. The pairings with the fewest drivers return the highest, including every match correct in the GT3 against road car comparison, where five drivers were available and chance is 0.20. Within individual circuits, where the number of people is smaller still, one of seven comparisons reaches significance in either population.

The signal in this section comes from the repeatability of the individual metrics rather than from the identification test. Of the 19 metrics that make up the fingerprint, 3 survive correction in the restricted population and 5 in the extended one. In both populations the strongest is apex speed, with an adjusted repeatability of 0.53 in the restricted population and 0.64 in the extended one, and its interval runs from 0.11 to 0.70 in the first and from 0.43 to 0.77 in the second. The two metrics that follow it in the restricted population are the spread of exit speed across corners, at 0.225, and the spread of apex speed across corners, at 0.224. The difference between the two, 0.001, is far smaller than the width of either interval. The interval of the exit speed spread starts at 0.016, while the interval of the apex speed spread reaches zero. Each of these intervals rests on 200 bootstrap resamples, a count that leaves the bounds themselves imprecise.

\begin{figure}[!htb]
\centering
\includegraphics[width=0.72\textwidth]{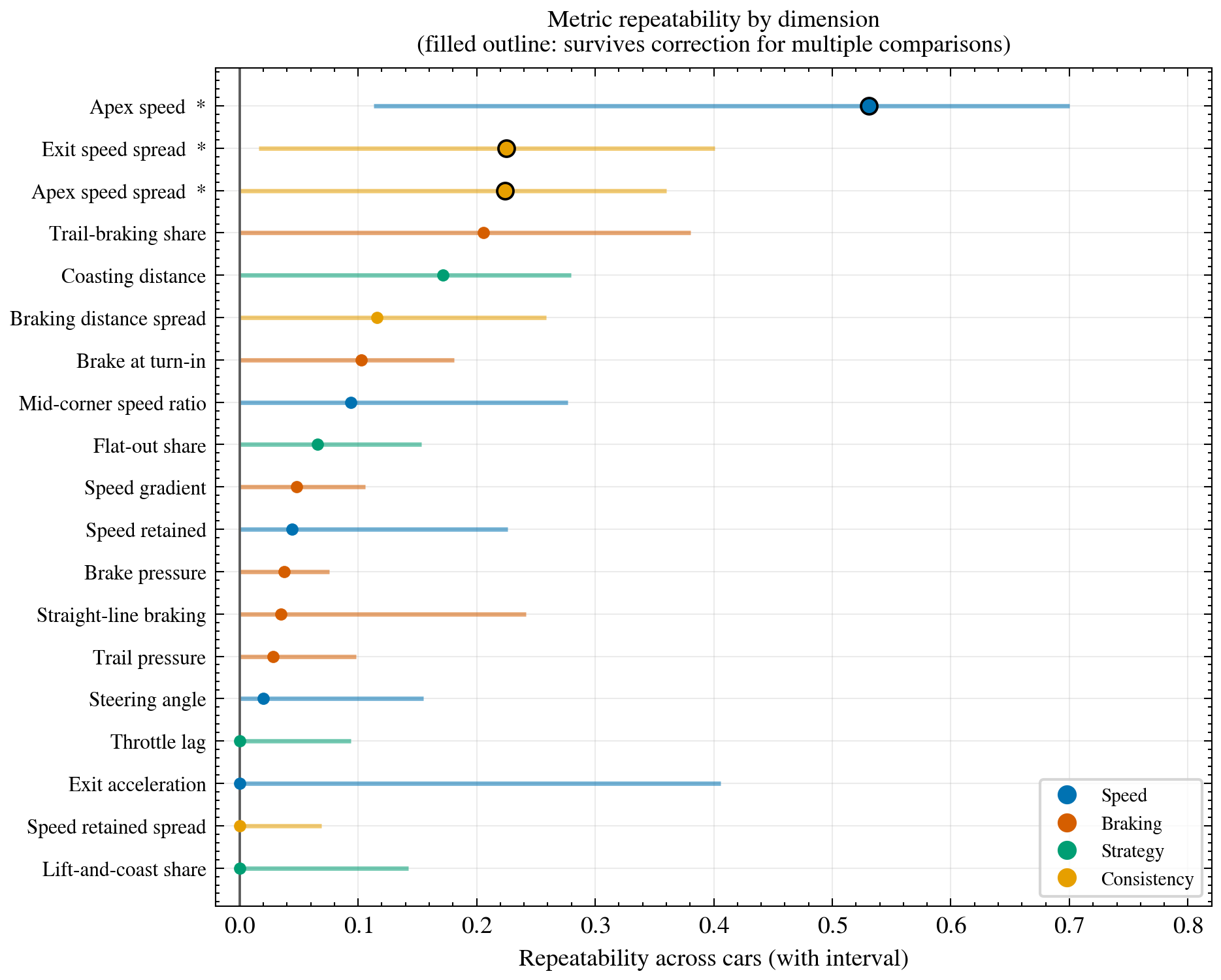}
\caption{Repeatability of each metric across cars in the restricted population, with its interval, colored by dimension.\label{fig:repeat}}
\end{figure}

One of the more informative findings here is how these metrics distribute across the four dimensions. In the restricted population, every surviving metric falls in either the speed or the consistency dimension, and none of the six braking metrics or the four strategy metrics survives correction. The extended population adds one metric in each of those two dimensions, trail-braking share at 0.17 and coasting distance at 0.24. Figure~\ref{fig:repeat} shows all 19 estimates in the restricted population, each with its interval and colored by its dimension. Both properties can be said to matter at once. In this sample, the dimension a metric falls in tracks whether it survives correction; the width of its interval sets how much weight the surviving value can carry. A property of the driver is therefore recoverable in the primary reading, but a narrow one: how fast a corner is taken, and how much that pace spreads across the corners of the circuit.

Two assumptions were registered before the data were examined, and two hypotheses were derived from them. Testing showed that neither was supported. The first held that dimensionless metrics, being free of the units in which a car expresses itself, would repeat across cars better than dimensional ones. Comparing the 5 dimensionless metrics against the 14 dimensional ones gives median repeatabilities of 0.066 and 0.043 in the restricted population and 0.078 and 0.039 in the extended one, and the difference does not reach conventional significance in either ($p = 0.55$ and $p = 0.41$). The second held that ratios derived from pairs of raw metrics would carry the driver more reliably than the raw metrics themselves. Three such ratios could be computed and two could not, and none of the three met the criterion registered for it. One of the three, however, repeats better than all but one of the raw metrics: the ratio of exit speed to apex speed returns 0.30 in the restricted population and 0.35 in the extended one, where its interval runs from 0.18 to 0.47.

The degree to which this design was underpowered for the question it asks can itself be quantified. Achieving an interval of width 0.30 around a repeatability of 0.5 requires 58 drivers, and around 0.6 requires 46. The restricted population provides 13 inferred drivers and the extended one 16, roughly a fifth to a third of what is needed.

This is visible in the results as well. The interval around the strongest repeatability estimate in the restricted population has a width of 0.59, against the 0.62 the calculation anticipates at that value.

Three aspects of the measurement procedure were corrected during the analysis, and their effect on the results did not run in the same direction. Estimating the whitening transform with the matched person left out, and applying the false discovery rate correction across car pairings rather than within them, both moved the identification result away from significance. By contrast, replacing the optimizer used to fit the mixed models moved the repeatability result toward it, raising the number of metrics surviving correction from two to three and reducing the number of boundary solutions from 15 to 3. All three are described in the preregistration amendment.

\begin{table}[htbp]
\footnotesize
\setlength{\tabcolsep}{4pt}
\renewcommand{\arraystretch}{1.15}
\caption{Cross-car identification by pairing. Accuracy is the proportion of drivers matched to their own record in the other car; $q$ is the false discovery rate-adjusted $p$-value across the three pairings within each population.\label{tab:crosscar}}
\begin{tabularx}{\textwidth}{@{}X l c c c c@{}}
\toprule
\textbf{Population} & \textbf{Pairing} & \textbf{$n$} & \textbf{Accuracy} & \textbf{Chance} & \textbf{$q$}\\
\midrule
Restricted & GT3 -- open-wheel & 12 & 0.25 & 0.08 & 0.191\\
Restricted & GT3 -- road car & 5 & 1.00 & 0.20 & 0.144\\
Restricted & Open-wheel -- road car & 4 & 0.50 & 0.25 & 0.191\\
\midrule
Extended & GT3 -- open-wheel & 15 & 0.33 & 0.07 & \textbf{0.022}\\
Extended & GT3 -- road car & 5 & 1.00 & 0.20 & \textbf{0.046}\\
Extended & Open-wheel -- road car & 4 & 0.50 & 0.25 & 0.074\\
\bottomrule
\end{tabularx}
\end{table}

\subsection{Normalization Against a Reinforcement-Learning Reference}
\label{sec:resref}

The reference was built from the reinforcement-learning agent sessions distributed with the dataset. For each circuit the first 150 clean laps were retained, and every lap was segmented with the same procedure applied to the human recordings, yielding 1650 corner passes at Monza, 2400 at Barcelona, and 1500 at the Red Bull Ring. All of them passed the validity criteria, so the reference carries every corner at every circuit. The 150-lap figure is a ceiling: the agent sessions available at each circuit exceed it. The retained laps come from a single car.

Reference behavior differs between circuits (Table~\ref{tab:reference}). Mean coasting distance ranges from 9.43~m at Monza to 1.80~m at the Red Bull Ring, a factor of 5.2, while mean apex speed follows a different ordering: Monza and the Red Bull Ring fall within 1~km/h of each other at 156.8 and 156.0~km/h, with Barcelona well below at 141.0~km/h. Mean brake pressure and trail-braking share also vary across circuits. A reference constructed by pooling circuits would therefore carry the circuit into the normalized value: the same human behavior would be scored differently at Monza than at the Red Bull Ring for reasons that have nothing to do with the driver.

Ten of the 19 metrics are expressed as ratios to the reference and nine retain population scaling; Table~\ref{tab:metrics} marks each. These ten sit unevenly across the dimensions: four in speed, four in braking, two in strategy, and none in consistency. A gap in that dimension is therefore undefined by construction.

\begin{figure}[!htb]
\centering
\includegraphics[width=0.62\textwidth]{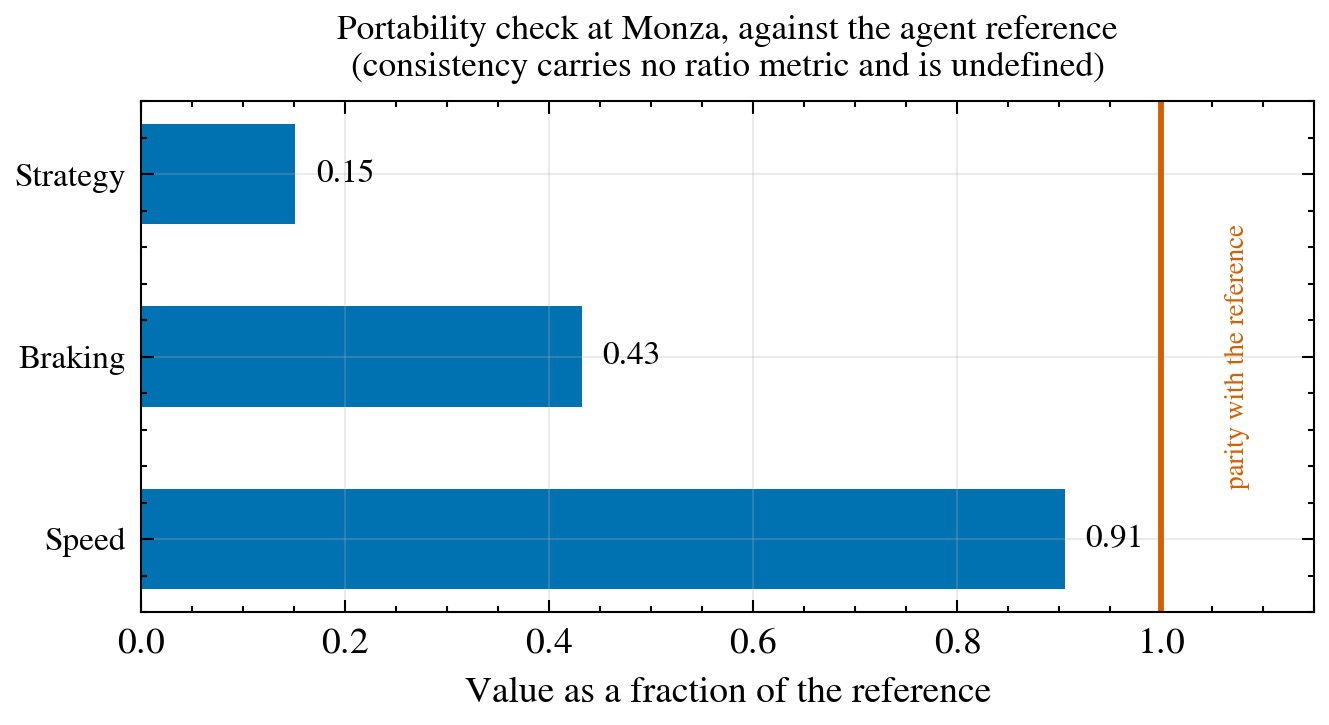}
\caption{Portability check at Monza. Each dimension is shown as a fraction of the agent reference; the consistency dimension carries no ratio metric and is undefined.\label{fig:gap}}
\end{figure}

As a portability check we applied the pipeline unchanged to the first author's sessions at Monza, recorded on a different simulator, in a different car class, and exported through a different telemetry tool. It produced a score without modification: 0.91 in the speed dimension, 0.43 in braking, and 0.15 in strategy, with consistency undefined for the reason just given. The values were identical under both population definitions, as expected, since the ratios depend only on the reference. Figure~\ref{fig:gap} sets the dimensions side by side against the reference. The three that are defined do not rest on equal evidence. The strategy dimension carries the value furthest from parity, and it is also the dimension built from the fewest ratio metrics, so it is the value that should be treated with the most caution.

Four limits bound what this section shows. The reference is a systematic subsample, since laps were taken in a fixed order rather than drawn at random, so it represents whatever portion of the released agent sessions that order reaches before the cap is met. Some of the ratios rest on small denominators, most sharply at the Red Bull Ring, where a mean coasting distance of 1.80~m turns small absolute differences into large ratios. In the strategy dimension one metric is degenerate at every circuit, taking the same value for every record. And the demonstration is cross-platform: the difference between the two simulators is part of what the reported values contain, so they measure the pipeline's reach and not the driver's skill.

\begin{table}[htbp]
\footnotesize
\setlength{\tabcolsep}{4pt}
\renewcommand{\arraystretch}{1.15}
\caption{The reinforcement-learning reference, built from the first 150 clean laps at each circuit.\label{tab:reference}}
\begin{tabularx}{\textwidth}{@{}X ccccc@{}}
\toprule
\textbf{Circuit} & \textbf{Corner passes} & \textbf{Apex speed (km/h)} & \textbf{Brake pressure} & \textbf{Trail share} & \textbf{Coasting (m)}\\
\midrule
Monza & 1650 & 156.76 & 0.1584 & 0.2267 & 9.428\\
Barcelona & 2400 & 140.95 & 0.1098 & 0.1925 & 4.141\\
Red Bull Ring & 1500 & 155.95 & 0.1132 & 0.2200 & 1.798\\
\bottomrule
\end{tabularx}
\end{table}
    \subsection{Comparison with a Lap-Time Baseline}
    \label{sec:resbase}

    The studies reviewed in Section~\ref{sec:rw-outcome} group laps by lap time and read driving behavior off the resulting groups. That grouping is available in this data as well, and it gives a baseline against which the unsupervised grouping used here can be placed. This subsection reports two measurements: how lap time distributes between and within recording sessions, and how far the grouping built from it agrees with the grouping built from the 19 metrics.

    Lap time is not part of the fingerprint and is not carried in the derived tables, so it was recovered from the processed telemetry. A lap was taken as complete when its normalized track position began below 0.05 and ended above 0.95. Of the 715 lap records, 603 satisfy that criterion, across 27 recording sessions and nine circuit and car cells. Because a lap at Monza is not comparable to a lap at the Red Bull Ring, every lap time was standardized within its own circuit and car cell before use. This recovery is independent of the derived tables, so its lap counts do not line up with those in Section~\ref{sec:respop}. Removing the excluded identifiers, and then any session left with fewer than two complete laps, leaves 476 laps across 19 sessions in the restricted population and 594 across 22 in the extended one. Of those sessions, 13 and 16 respectively also carry a fingerprint, and those are the ones that enter the comparison below.

    The first measurement is the share of lap-time variance that separates sessions from one another. In the restricted population that share is 0.33, with an interval running from 0.14 to 0.46, and in the extended population it is 0.41, from 0.24 to 0.52. Two independent estimators, a mixed model and a one-way analysis of variance, agree to within 0.012, and both intervals rest on 200 bootstrap resamples of the recording sessions. Roughly three fifths to two thirds of the variance in lap time therefore falls within sessions rather than between them. In this data lap time is a property of the lap before it is a property of the driver.

    The second measurement compares the two groupings directly. Sessions were placed in three lap-time groups by their median standardized lap time, matching the three clusters the fingerprint returns, and the two partitions were compared with the adjusted Rand index against a permutation null of 10,000 draws. The index is 0.48 in the restricted population and 0.47 in the extended one, and it reaches significance in both ($p = 0.005$ and $p = 0.001$). Of the results this paper reports, this is the only one that does not change direction when the population definition changes.

\begin{figure}[!htb]
    \centering
    \includegraphics[width=\textwidth]{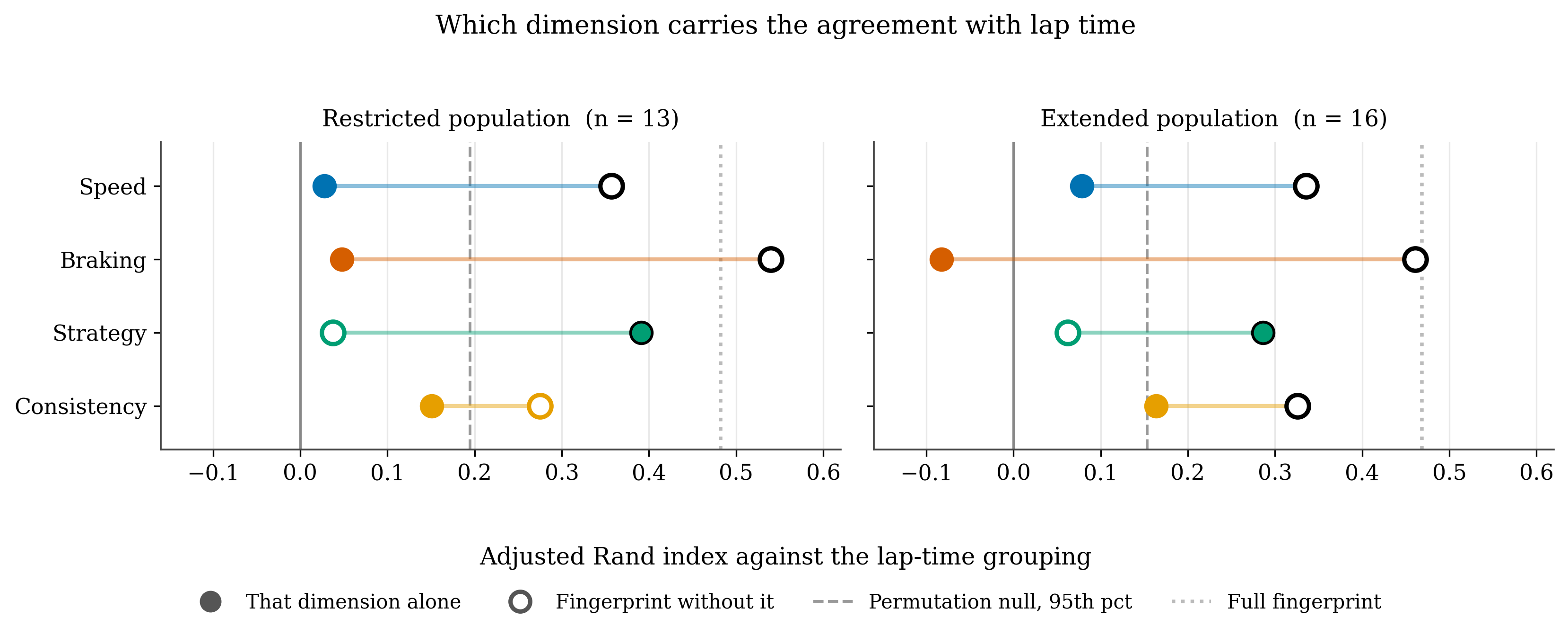}
    \caption{Agreement with the lap-time grouping, by dimension. Left: the restricted
    population; right: the extended one. A filled marker is the clustering built from
    that dimension alone, an open marker the full fingerprint with that dimension
    removed, and a black outline a permutation value below 0.05.\label{fig:aridim}}
    \end{figure}

    Because lap time and pace are closely related, the agreement invites a simple explanation: that the fingerprint recovers the speed dimension and little else. Testing does not support it. Clustering on the speed dimension alone returns an index of 0.03 in the restricted population and 0.08 in the extended one, and neither reaches significance. Removing that dimension from the fingerprint lowers the index to 0.36 and 0.34, where it remains significant. The dimension that carries the agreement is strategy (Figure~\ref{fig:aridim}). Clustering on it alone returns 0.39 and 0.29, the only single dimension to reach significance in either population, and removing it drops the index to 0.04 and 0.06, where it does not. The clustering that remains once strategy is removed is not degenerate: its groups hold six, five and two records in the restricted population and seven, five and four in the extended one, and its silhouette value, at 0.285 and 0.355, stays within 0.03 of the value the full fingerprint returns and rises above it in the extended population. What disappears when the strategy dimension is removed is the relation to lap time, and not the grouping itself.

    Three limits bound this comparison. The agreement rests on 13 records in the restricted population and 16 in the extended one, so the index is estimated on a small sample even where its null distribution is not in doubt. The lap-time grouping divides sessions into three, following the number of clusters the fingerprint returns rather than the six levels used in the study the baseline is drawn from. Repeating the comparison with two through six lap-time groups leaves the agreement above chance at every number in the extended population, and at two, three and six in the restricted one, where four and five fall back to the null. And one observation is left unexplained: the speed dimension correlates with median lap time at least as strongly as the strategy dimension does, at $-0.62$ against $0.64$ in the restricted population, yet it produces no agreement at all when the grouping is built from it.

    \section{Discussion}
    \label{sec:discussion}

    These three tests run independently of one another, and they ask different things. The first asks whether a driver's grouping carries from one corner type to another; the second, whether the metrics repeat when the car changes; the third, what happens to the separation between clusters as the range of available telemetry widens. All three answer in the same way. In the restricted population the grouping does not carry from one corner type to the next, and in the two pairings that involve slow corners the agreement falls below chance; when the car changes, only three of the 19 metrics survive correction; and as coverage widens, the excess of the separation over its null first narrows and then changes sign. What the three have in common is that the structure we recover is conditional on which corner, which car, and which coverage we look at. That it is a property of the driver is not shown by these data.

    This reading holds for the restricted population. In the extended population, two of the three car pairings survive correction, and the two narrower rungs of the ladder separate from their null. In short, the same pipeline returns a partly different answer when the sample is defined more broadly. This sensitivity is not a defect but a part of the result: some of what carries the finding is how the population is drawn, and not the drivers themselves. Of the four results, the only one that transferred without modification was the reference normalization, and what transferred is a measurement device. The pipeline works; what it measures, however, is less specific to the driver than the framing assumed at the outset.

    Along two axes, we tested whether grouping reflects the driver more closely as more data is pooled, holding the number of dimensions constant so values along each axis remain comparable.

    Across the tier ladder, where each of the first three steps adds a car class, the silhouette value of the three-cluster solution falls at every step and never rises: 0.377, 0.300, and 0.265 in the restricted population. The fourth step adds a circuit, and its values repeat the third, because that circuit contributes no processed human sessions.

    Along the second axis, adding three records to the same feature space improves every measure. The silhouette value rises from 0.188 to 0.211, the mean agreement between the within-type clusterings rises from 0.043 to 0.271, and the permutation value falls from 0.119 to 0.001, the floor of the test.

    The three records that produce this improvement are the three whose person-level identity the design does not establish.

    Two readings of the pattern remain open. Under the first, the three records come from three distinct drivers, and the wider definition recovers a signal that the narrower one is too small to show. Under the second, two or more of them come from the same driver, and agreement rises because one person is counted as several. The design distinguishes neither.

    Once per car pairing, we tested whether the pipeline separates drivers rather than sessions, and we calibrated the false-positive rate before reading any of the three tests.

    In the restricted population, all three tests fail the multiple-comparison correction, the adjusted values being 0.191, 0.144, and 0.191; in the extended population, two of the three pass, at 0.022 and 0.046, with the third reaching 0.074.

    Of the 19 metrics, three show a repeatability that remains after the same correction is applied. The strongest of them is the mean apex speed, having a repeatability coefficient of 0.531 and a confidence interval ranging from 0.113 to 0.700. The interval is wide, so the coefficient establishes that a driver effect is present without establishing its size.

    The calibration test is the only instance in which the design has an experimental character. When the whitening transform is fitted on the matched pairs and then tested on them, the false-positive rate amounts to 96.7 percent over 30 trials. When the same transform is fitted with the tested pair left out, the rate drops to 6.7 percent against a nominal five. The procedure that withholds the tested pair is therefore calibrated. The one that does not is not. The gap between these two rates, rather than either rate on its own, is what 30 trials support.

    The day-to-day rung of the identification ladder carries four individuals in its largest cell. At that size a perfect match returns a value of 0.042, and the next attainable value is 0.29. The rung thus distinguishes between two states, and it provides no measure for the values in between.

    A null result at this sample size is a statement about the design. What the telemetry contains is left where it was.

    The analysis plan was written before the measurements were taken, and each departure from it was recorded with its reason. One of those departures changed the optimizer used to fit the repeatability models, and it moved a result toward significance. It is recorded in the same list as the others.

    Two hypotheses were registered in advance, and neither held. Metrics without physical units were expected to transfer across cars more stably than metrics with them; the most stable metric in the set carries units. No derived ratio reached the threshold of 0.40 the plan had set, and two of the ratios could not be computed at all, because the columns they require are absent from the source data.

    The pipeline that produced these measurements runs end to end. It reads two telemetry formats, segments three circuits into 37 corners, derives 19 metrics, and carries them through clustering, identification testing, and normalization against a reinforcement-learning reference.

    Every limit reported here is a measurement the pipeline produced.


    \section{Conclusions}
    \label{sec:conc}

    The analysis plan described a three-rung ladder, and the first rung was never implemented. Within-day repeatability, which would have set the ceiling for the two rungs above it, was not measured. Every coefficient reported here is therefore read against a ceiling whose height is unknown.

    The requirements listed above each raise what the design can detect. None of them separates how much of the remaining gap belongs to the driver and how much to the measurement. That separation is what the missing rung would supply, and it can be estimated from the recordings already collected, by splitting the laps within a single session. It is the cheapest of the requirements, and it is the one the others are read against.

    \vspace{6pt} 
    

    \section{Future Works}
    \label{sec:futurework}
    
    The clearest requirement is sample size. The power calculation in Section~\ref{sec:rescars} puts the number
    of drivers needed for an interval of width 0.30 around a repeatability of 0.5 at 58, against the 13 the
    restricted population provides; until that gap closes, the cross-car question is left open. Two further
    limits are structural. The person-level grouping used there is inferred from the record identifiers and is
    confirmed by no label in the data, so a corpus that records driver identity directly would settle what the
    present design can only condition on. And because the clusters are built from the same 19 metrics the
    surrogate then predicts them from, the attribution reads the geometry of the clustering rather than an
    independent fact about driving; an external criterion, recorded separately from those metrics, is what
    would break that circularity.
    
    Two components stand between the framework and the coaching system it is named for. The first is validation
    against known ground truth: the corner inventory here matches the published layouts by construction, so a
    controlled environment in which the geometry is specified in advance would test the detector instead of
    confirming it. The second is the step from attribution to advice. The attributions report which metrics
    separate the groups; they do not give a driver a value to aim for, and a recommendation needs a threshold,
    which a ranking does not supply. A prospective study, in which drivers receive the output and are measured
    again afterwards, is what would show whether the advice changes anything.

    In the future,we would like to implement our TRACK on mobile platforms. We believe that our framework will work on mobile platforms at real-time frame rates. Additionally, there is potential for upgrading and optimizing our TRACK by incorporating state of the art deep learning methods and techniques~\cite{Gok2023SIU, Azadvatan2024arXiv, Jabbarli2024arXiv,Akdogan2025ARTEMIS,Turkun2026MDPID,Akdogan2026MDPIM,Azadvatan2026MDPIAS}.

    To improve accuracy of our TRACK, we are also interested in implementing realistic Bidirectional Reflectance Distribution Function (BRDF)~\cite{Ozturk2006EGUK, Kurt2007MScThesis, Ozturk2008CG, Kurt2008SIGGRAPHCG, Kurt2009SIGGRAPHCG, Kurt2010SIGGRAPHCG, Ozturk2010GraphiCon, Szecsi2010SCCG, Ozturk2010CGF, Bigili2011CGF, Bilgili2012SCCG, Ergun2012SCCG, Toral2014SIU, Tongbuasirilai2017ICCVW, Kurt2019DEU, Tongbuasirilai2020TVC, Akleman2024arXiv,Kurt2025arXiv}, Bidirectional Scattering Distribution Function (BSDF)~\cite{WKB12, Ward2014MAM, Kurt2014WLRS, Kurt2016SIGGRAPH, Kurt2017MAM, Kurt2018DEU}, Bidirectional Surface Scattering Reflectance Distribution Function (BSSRDF)~\cite{Kurt2013TPCG, Kurt2013EGSR, Kurt2014PhDThesis, Onel2019PL, Kurt2020MAM, Kurt2021TVC, Yildirim2026MDPIAS} and multi-layered material~\cite{WKB12, Kurt2016SIGGRAPH, Mir2022DEU} models into our framework (TRACK).
    
    
    
    
    \authorcontributions{Conceptualization, E.Ç. and M.K.; methodology, M.K.; software, E.Ç.; validation,  E.Ç. and M.K.; formal analysis,  E.Ç. and M.K.; investigation,  E.Ç. and M.K.; resources,  E.Ç. and M.K.; data curation, E.Ç.; writing---original draft preparation, E.Ç.; writing---review and editing,  E.Ç. and M.K.; visualization, E.Ç.; supervision, M.K.; project administration, M.K. All authors have read and agreed to the published version of the manuscript.}

    \funding{This research received no external funding.}

\dataavailability{
The analysis code that produced the results reported in this paper is held in the TRACK
repository on GitHub at \url{https://github.com/EfeCangirili/TRACK-sim-racing-telemetry}.
The repository records how each notebook is run and which section of this paper it feeds.
along with the derived tables and the figures this paper reports.
The primary telemetry analyzed here is the Assetto Corsa Gym (ACGym) dataset of
Remonda et al.~\citep{remonda2024acgym}, distributed through Hugging Face as
\href{https://huggingface.co/datasets/dasgringuen/assettoCorsaGym}{\texttt{dasgringuen/assettoCorsaGym}} under the Creative Commons Attribution 4.0
International license (\url{https://creativecommons.org/licenses/by/4.0/}). Revision
\texttt{ad1df12c49cce10f159ba31967744750bb9edd8b} was used. The dataset is not
redistributed here. It was filtered by the validity criteria of Section~\ref{sec:data},
divided into corner passes by the procedure of Section~\ref{sec:corners}, and reduced to
the derived metrics of Section~\ref{sec:fingerprint}; those are the changes made to it.
The Assetto Corsa Competizione sessions recorded by the first author and used for the
portability check in Section~\ref{sec:resref} are not part of ACGym. They are archived
separately and are available from the first author on request.}

    \acknowledgments{The authors thank Remonda et
    al.~\citep{remonda2024acgym} for releasing the Assetto Corsa Gym dataset under
    an open license, which made this analysis possible.}
    
    \conflictsofinterest{The authors declare no conflicts of interest.} 

    \begin{adjustwidth}{-\extralength}{0cm}
    
    \reftitle{References}
    
    
    \bibliography{MDPI26_TRACK_References_V1}

    %
    
    
    \PublishersNote{}
    \end{adjustwidth}
    \end{document}